\documentclass[aps, prd, preprint, groupedaddress, nofootinbib]{revtex4-1}

\usepackage{amssymb,amsmath,graphicx,latexsym}
\usepackage{xcolor}

\usepackage{subfigure}

\usepackage{multirow}

\usepackage{graphicx}
\usepackage{epsfig}
\usepackage{bm}
\usepackage{amsfonts}
\usepackage{amsmath}
\usepackage{amssymb}
\usepackage{multirow}
\usepackage{subfigure}
\usepackage{wasysym}
\usepackage{array}
\usepackage{diagbox}
\usepackage{makecell}
\usepackage{slashbox}
\usepackage[pdftex,hidelinks]{hyperref}
\usepackage[T1]{fontenc}
\usepackage[utf8]{inputenc}
\usepackage{array}
\usepackage{multirow}
\usepackage{amsmath}
\usepackage{amssymb}
\usepackage{stackrel}
\usepackage{rotating}
\usepackage{esint}

\begin{document}



\title{Cosmological perturbation theory with trinity of scalar fields}





\author{
Amjad Ashoorioon$^{1}$\footnote{amjad@ipm.ir}
}

\author{
Shinji Mukohyama$^{2}$\footnote{shinji.mukohyama@yukawa.kyoto-u.ac.jp}
}

\author{
Kazem Rezazadeh$^{3}$\footnote{kazem.rezazadeh@ipm.ir}
}

\author{
Navid Talebizadeh$^{4}$\footnote{navidtalebizadeh@ipm.ir}
}
\affiliation{$^{a,c,d}$School of Physics, The Institute for Research in Fundamental Sciences (IPM)\\P.O. Box 19395-5531, Tehran, Iran}

\affiliation{$^b$Center for Gravitational Physics and Quantum Information, Yukawa Institute for Theoretical Physics,\\
Kyoto University, 606-8502, Kyoto, Japan}
\affiliation{$^b$Research Center for the Early Universe (RESCEU), Graduate School of
Science, The University of Tokyo, Hongo 7-3-1, Bunkyo-ku, Tokyo
113-0033, Japan}
\affiliation{$^b$ Kavli Institute for the Physics and Mathematics of the Universe (WPI),\\
The University of Tokyo Institutes for Advanced Study, The University of Tokyo, Kashiwa, Chiba 277-8583, Japan}



\date{\today}


\begin{abstract}
We present an explicit formulation of cosmological perturbation theory for three-field models with a flat field space. By performing rotations to align one field with the direction of curvature perturbations and applying the same rotations to the other two field directions, we introduce the semikinematic basis, which is applicable to models with more than two fields. We derive the governing equations in this basis. We also stress a characteristic property of more-than-two-field models: the freedom in choosing the isocurvature perturbations. This framework enables the computation of the curvature and two isocurvature power spectra for any given potential. We numerically solve the background and perturbation equations for three distinct scenarios. First, to validate the consistency of our three-field formalism, we examine an effective two-field model inspired by the two-block case of the multigiant vacua matrix inflation scenario. Next, we analyze a purely three-field system without direct interfield interactions. Finally, we study a three-field case that incorporates direct interactions. For all scenarios, we numerically compute the curvature perturbation power spectra and highlight the effects of rapid turns on the spectra. Finally, we investigate the relationship between these quantities and the observables in the early radiation-dominated era. Through both general arguments and a simple example, we show that three-field inflation can yield a much richer phenomenology. This is particularly true when we assume the initial perturbations in the radiation era include two isocurvature modes.

\end{abstract}



\maketitle



\section{Introduction}
\label{sec1}

The inflationary paradigm has become an essential part of modern cosmology, providing a compelling explanation for the observed large-scale homogeneity and isotropy of the Universe, and also a mechanism for explaining the origin of the small anisotropies observed in the temperature fluctuations in the cosmic microwave background (CMB). Inflation also provides phenomenological predictions that could be tested by current and future experiments. For instance, the fact that inflation produces a red scalar power spectrum is expected from the fact that inflation has ended. This is compatible with the latest result of the Planck 2018 experiment which determines $n_s=0.9649\pm 0.0042$ \cite{Planck:2018jri}. Inflationary models naturally produce gravitational waves (GWs), whose amplitude, however, depends on the scale of inflation, and hence, may be unobservable if the scale of inflation is too low. Although inflation was not designed to explain the origin of primordial black holes (PBHs), they can be used to explain the fluctuating modes with large amplitudes needed to explain the origin of PBHs. An inflationary epoch can also act like a very high-energy lab that can provide us with deeper insights about the species with masses around the inflationary Hubble parameter \cite{Arkani-Hamed:2015bza}.

Inflation is a paradigm and there are a plethora of models that can give the desired behavior for the early evolution of the Universe. The simplest models of inflation are the single-field models. They successfully fulfill the current observational constraints, most of which come from the CMB data \cite{WMAP:2008lyn,Planck:2018jri}. The single scalar degree of freedom (DoF) during inflation can be explained from the effective field theory point of view \cite{Cheung:2007st, Ashoorioon:2018uey}, in which one expects a Goldstone boson from the breaking of the time translation symmetry \cite{Arkani-Hamed:2003pdi,Creminelli:2006xe}. Nonetheless, there are some phenomenological and theoretical motivations to think about more complicated models that involve more than one field.

Single-field models predict the spectrum of perturbations to be nearly Gaussian, $f_{{}_{\rm NL}}^{\rm loc} \sim \mathcal{O}(0.01)$. However, current available observations do not exclude the possibility of having larger non-Gaussianities, e.g. from the Planck 2018 experiment \cite{Planck:2019kim}, we have $f_{{}_{\rm NL}}^{\rm loc}= -0.9 \pm 5.1$. Future observations may severely constrain the non-Gaussianity \cite{Dore:2014cca,Abazajian:2019eic}. Multifield models of inflation are one of the candidates that can produce non-Gaussianities in the primordial spectra. Another phenomenological motivation is related to the enhancement of the scalar power spectrum on scales smaller than the observed scales, e.g. in CMB \cite{Ivanov:1994pa,Garcia-Bellido:2017mdw,Ashoorioon:2019xqc}. These enhancements can in turn lead to other observable phenomena, e.g. PBH production \cite{Carr:1974nx}, which may fully or partly contribute to the dark matter content of the Universe \cite{Bartolo:2018evs,Bertone:2018krk,Bird:2016dcv}, and also the production of the second-order GWs \cite{Ananda:2006af,Baumann:2007zm}. Some mechanisms have been introduced that can enhance the scalar power spectrum on relevant scales e.g. excited states \cite{Fumagalli:2021mpc}, resonant amplification \cite{Peng:2021zon}, having different dispersion relations\cite{Ashoorioon:2019xqc}, and phase transition\cite{An:2020fff}. Another possible mechanism to enhance the power spectrum is to have sharp turns in the field trajectory in multifield inflationary models (see e.g. \cite{Achucarro:2010da,Konieczka:2014zja}). Having turns in the trajectory of the field space, also known as the nongeodesic trajectory in the literature, can also cause some other interesting phenomena, e.g. generation of derivative interactions between the adiabatic and nonadiabatic modes \cite{GrootNibbelink:2001qt,Gordon:2000hv,GrootNibbelink:2000vx}, and particle production due to the nongravitational interactions arising from the nongeodesic motion in the field space \cite{Parra:2024usv}.

Multiple-field models of inflation also have some dynamical properties that are absent in the single-field models, the most significant of which is the appearance of isocurvature perturbations. The presence of isocurvature modes can alter the overall curvature perturbation during inflation, which in turn can make a detectable non-Gaussianity. There would also be some residual isocurvature fluctuation which can be correlated to the curvature perturbations \cite{Bassett:2005xm}. In this sense, detection of primordial isocurvature perturbations would be a sign for more complicated models of inflation. However, there are some observational constraints on the isocurvature modes in the early radiation-dominated era, the most important of which arises from CMB anisotropies. The structure of acoustic peaks for the pure isocurvature, and pure adiabatic initial condition of perturbation is a very distinctive feature. Detection of the first acoustic peak at $l \simeq 220$ ruled out a pure isocurvature mode as a sole source of perturbations \cite{Enqvist:2001fu}. However, presence of a subdominant isocurvature contribution is not still ruled out. Degrees of freedom other than the inflaton can also produce the primordial perturbations, the most well-known scenario of which is the curvaton model \cite{Lyth:2001nq}.
There are also some theoretical motivations for inflationary models with more than one scalar field. From the perspective of UV physics, existing multiple scalar fields is a natural expectation. In supergravity and stringy models there exists a plethora of scalar moduli fields(see e.g. \cite{Kachru:2003sx,Firouzjahi:2003zy,Baumann:2014nda,Baumann:2007ah,Burgess:2006nod,McAllister:2007bg,Cline:2006hu,HenryTye:2006uv,Ashoorioon:2006wc,Ashoorioon:2008qr,Ashoorioon:2009sr,Ashoorioon:2009wa,Ashoorioon:2011ki}), and in theories that gravity propagates in extra dimensions, infinite series of particles arise from the extra dimension and the inflaton field can mix with these fields through the coupling to the higher-dimensional Ricci scalar. Another theoretical motivation comes from the swampland conjectures \cite{Obied:2018sgi,Agrawal:2018own,Garg:2018reu,Ooguri:2018wrx}, which although still under debate, put constraints on the parameters of low-energy models that possess a UV completion in string theory, and from these constraints having multiple fields in the inflationary model is motivated \cite{Achucarro:2018vey}.\\
The dynamics of different DoFs during the inflation depend on their masses. The lightness and heaviness of the modes in this context are determined compared to the Hubble parameter. Therefore, typically we have three classes of modes which are light modes, $m < H$, heavy modes, $m > H$, and modes that have masses at the order of Hubble parameters, $m \sim H$, which we call medium mass modes. Usually the light modes are the ones that drive inflation. Heavy modes usually do not contribute to the inflationary dynamics, since the amplitude of their quantum perturbations decays exponentially after exiting the horizon. Effects of existing medium mass modes besides the light inflaton mode is investigated in the quasisingle-field model \cite{Chen:2009zp}. Having three types of DoFs according to their masses shows that a three-field model can be a typical case that can potentially include all different types of modes. In addition, three-field inflationary models have also some unique properties, different from the two-field case, and more similar to more-than-three-field models, enabling a more direct generalization to $N$-field models with $N > 3$ \cite{Aragam:2023adu,Aragam:2024nej}. One of these properties is related to the characteristic of isocurvature perturbation. In two-field models, the isocurvature perturbation is defined as the perturbations orthogonal to the adiabatic direction in the field space which is a unique direction in two-dimensional field space. However, in more-then-two field models, there is a hypersurface orthogonal to the adiabatic direction which gives us a freedom to choose the isocurvature direction. This issue is investigated in this paper.

 So far, several more-than-one-field inflationary scenarios have been proposed, most of which are two-field models, e.g. assisted inflation models \cite{Liddle:1998jc}, N-flation \cite{Dimopoulos:2005ac}, hybrid inflation \cite{Linde:1993cn}, and curvaton models \cite{Lyth:2001nq}. A general multiple-field model has also been introduced in
\cite{Ashoorioon:2009wa,Ashoorioon:2009sr}, which is called matrix-inflation (M-flation). In this model, which is inspired by string theory and D-brane dynamics, the inflaton fields were taken to be matrix-valued objects. It is shown that this model can be reduced to a standard single- or multiple-field model at the classical level, however depending on the matrix dimension, there could be many isocurvature modes with a specific mass spectrum at the quantum level. General formalisms of multifield dynamics, in both the background and perturbation level, have also been developed \cite{Wands:2007bd,GrootNibbelink:2000vx,GrootNibbelink:2001qt,Langlois:2008mn}, and the two-field case has been studied carefully \cite{Byrnes:2006fr,Lalak:2007vi}. Models with more than two fields have also been studied from various perspectives, such as in the context of the effective field theory of inflation \cite{Cespedes:2013rda} or with a focus on non-Gaussian features \cite{Pinol:2020kvw}. Additionally, the production of a stochastic gravitational wave background in three-field models was investigated in \cite{Aragam:2023adu,Aragam:2024nej}. However, unlike the two-field case—where explicit equations governing the adiabatic and isocurvature modes were derived for a given potential \cite{Lalak:2007vi}—such equations have not yet been obtained for the three-field scenario in these works.
\\
In this paper, we explicitly develop the three-field scenario and investigate both the background and perturbative regimes by numerically solving the equations for various cases, with a focus on the effects of rapid turns 
in the field trajectory. We perform a transformation from the field basis to the so-called semikinematic basis, which captures the adiabatic and isocurvature perturbations. We also explain the freedom in choosing this basis. 
Using this explicit formulation, we not only derive the spectrum of the curvature perturbation but also compute the evolution of isocurvature modes and their correlation with the curvature perturbation. For simplicity, we do 
not consider the curved field space in this work. We plan to investigate the nonflat field space case in the future. In Sec. \ref{sec2}, we construct the basic formulation of the background dynamics. We develop the 
perturbative regime in Sec. \ref{sec3}. We introduce the semikinematic basis and derive the governing equations for both adiabatic and isocurvature modes within this framework.  We solved the background and 
perturbative equations for three different cases numerically, and the results are shown in Sec. \ref{sec4}. The first case is an effectively two-field model, inspired by the two-block M-flation scenario. The second case is 
a completely three-field case, in which the fields have only self-interactions. The final case is a three-field case where the fields interact with each other, too. The effect of the turns in the background field trajectory is 
highlighted in the three-field cases. After completely developing the three-field formalism, we investigate in Sec. \ref{RTO} the relation between the inflationary correlation functions and the observable 
correlation functions in the early radiation-dominated era.  Finally, we conclude the paper in Sec. \ref{sec5}.

\section{Three-Field Cosmological Perturbation Theory: Background Level}
\label{sec2}

In this section, we are going to develop the three-field cosmological perturbation theory. As was mentioned in the Introduction, in the construction of inflation within more fundamental theories such as string theory, usually numerous fields are involved. Having multiple DoFs during inflation can have observational effects, which may be tested in the light of the upcoming experiments observing GWs or CMB at smaller scales. In this sense, investigating the three-field case explicitly would be a small but important step forward in understanding the possible multiple-field dynamics of the inflationary era.

For simplicity, we consider the simplest scenario, consisting of three scalar fields with canonical kinetic terms, \footnote{Generally, kinetic terms of the fields can be characterized by a metric for field space. As an example, the field space metric shows up when the fields are nonminimally coupled to the gravitational sector. By a conformal transformation, the coupling will become minimal again but the kinetic terms will be noncanonical and can be described by a metric on the field space \cite{Kaiser:2010ps}. Considering the UV-complete theory of the inflation, for example in the string theory framework, would also induce a nonflat metric for the field space.}
\begin{equation}\label{action}
    S = \int d^4 x\sqrt{-g} \left[\frac{1}{2}(\partial_{\mu}\phi)^2 + \frac{1}{2}(\partial_{\mu}\chi)^2+\frac{1}{2}(\partial_{\mu}\sigma)^2 +V(\phi , \chi , \sigma
) \right]\,.
\end{equation}
 The background equations of motion are
\begin{align}
     \Ddot{\phi} &+ 3H\Dot{\phi} + V_{, \phi} =0\,, \label{backequphi}\\
     \Ddot{\chi} &+ 3H\Dot{\chi} + V_{, \chi}=0\,,  \label{backequchi} \\
      \Ddot{\sigma} &+ 3H\Dot{\sigma} + V_{, \sigma}=0\,,  \label{backequsig} \\
      H^2 &= \frac{1}{3} \Big[  V + \frac{1}{2}(\Dot{\phi}^2 + \Dot{\chi}^2 + \Dot{\sigma}^2)  \Big]  \label{backequH}\,,\\ \nonumber
\end{align}
 in which a dot denotes the derivative with respect to the cosmic time. We set the reduced Planck mass, $M_{P}^2 \equiv \frac{1}{8\pi G}$ to 1. By solving these coupled equations, the background trajectory in the field space is determined. We are going to solve these coupled equations numerically with respect to $N_e$, the number of e-folds before the end of inflation. Our convention for the number of e-folds is $dN_e = Hdt$. We will follow the evolution from 80 e-folds before the end of inflation, $N_e = -80$, to the end of inflation at $N_e = 0$, and show the results for the last 60 e-folds from $N_e = -60$ to $N_e = 0$.

For an arbitrary quantity, $A$, we have the relation
\begin{equation}\label{transtN}
     \dot{A} = H A',
 \end{equation}
 where prime is the derivative with respect to $N_e$. The background Eqs. (\ref{backequphi})—(\ref{backequH}) with respect to $N_e$ will be
\begin{align}
   \phi'' &+ (3+\frac{H'}{H})\phi' + \frac{V_{, \phi}}{H^2} = 0\,, \label{backequphiN} \\
    \chi'' &+ (3+\frac{H'}{H})\chi' + \frac{V_{, \chi}}{H^2} = 0\,,  \label{backequchiN} \\
      \sigma'' &+ (3+\frac{H'}{H})\sigma' + \frac{V_{, \sigma}}{H^2} = 0\,, \label{backequsigN} \\
      H^2 &= \frac{1}{3} \Big[  V + \frac{H^2}{2}(\phi'^2 + \chi'^2 + \sigma'^2)  \Big]\,. \label{backequHN} \\ \nonumber
\end{align}

\section{Three-Field Cosmological Perturbation Theory: Perturbative Level}\label{sec3}

After solving the background equations and having the background trajectory, we are now going to discuss the linear perturbation regime of the three-field model. In \cite{Cespedes:2013rda}, the authors have developed the perturbative level of a three-field case by introducing a Goldstone boson as the fluctuation along the direction of broken time translational symmetry. They adopted the ADM metric and wrote the full action. Then by applying the constraint equations and considering the decoupling limit, equations of motion for the fluctuations at the linear regime were obtained. In this paper, however, we obtain the equations of motion for the Mukhanov-Sasaki variables at the linear regime without considering the decoupling limit. We follow \cite{Lalak:2007vi} and \cite{Gordon:2000hv}, and generalize them to the three-field case. Then, by solving the equations numerically, the desired observables will be calculated. As we will explain, the three-field case has some special features which are characteristic of more-than-two-field models.

For simplicity, we work in the longitudinal gauge and in the absence of anisotropic stress. \footnote{Although for more than one scalar field the energy-momentum tensor is not of the perfect fluid form to all orders in perturbations, it can be shown that to first order the anisotropic inertia vanishes (see \cite{Weinberg:2008zzc} p.499).} The perturbed metric is
\begin{equation}\label{pertmetric}
ds^2=-(1+2\Phi)dt^2+a^2(t)(1-2\Phi) d\vec{x}^2\,.
\end{equation}
Perturbations of the fields are
\begin{align}
\phi(t) \rightarrow \phi(t) + \delta \phi(\Vec{x},t)\,, \label{pertphi}\\
\chi(t) \rightarrow \chi(t) + \delta \chi(\Vec{x},t)\,, \label{pertchi}\\
\sigma(t) \rightarrow \sigma(t) + \delta \sigma(\Vec{x},t)\,. \label{pertsigma}\\ \nonumber
\end{align}
We are interested in the perturbation equations of Mukhanov-Sasaki variables, defined as
\begin{align}
Q_{\phi}=\delta \phi+\frac{\dot{\phi}}{H}\Phi\,, \label{MSphi}\\
Q_{\chi}=\delta \chi+\frac{\dot{\chi}}{H}\Phi\,, \label{Mschi}\\
Q_{\sigma}=\delta \sigma+\frac{\dot{\sigma}}{H}\Phi\,. \label{Mssigma}\\ \nonumber
\end{align}
Working in the Fourier space, and from Eqs \eqref{backequphi}—\eqref{backequsig}, Mukhanov-Sasaki variables satisfy the following linear order equations: 
\begin{align}
\ddot{Q_{\phi}}+3H\dot{Q_{\phi}}+\frac{k^2}{a^2}Q_{\phi}+\big[V_{,\phi \phi}-\frac{1}{a^3M_p^2}{\left(\frac{a^3}{H}\dot{\phi}^2\right)}^.\big]Q_{\phi}&+\big[V_{,\phi \chi}-\frac{1}{a^3M_p^2}{\left(\frac{a^3}{H}\dot{\phi}\dot{\chi}\right)}^.\big]Q_{\chi}\nonumber\\
&+\big[V_{,\phi \sigma}-\frac{1}{a^3M_p^2}{\left(\frac{a^3} {H}\dot{\phi}\dot{\sigma}\right)}^.\big]Q_{\sigma}=0\,, \label{diffequphi} \\
\ddot{Q_{\chi}}+3H\dot{Q_{\chi}}+\frac{k^2}{a^2}Q_{\chi}+\big[V_{,\chi \chi}-\frac{1}{a^3M_p^2}{\left(\frac{a^3}{H}\dot{\chi}^2\right)}^.\big]Q_{\chi}
&+\big[V_{,\chi \phi}-\frac{1}{a^3M_p^2}{\left(\frac{a^3}{H}\dot{\chi}\dot{\phi}\right)}^.\big]Q_{\phi}\nonumber\\
&+\big[V_{,\chi \sigma}-\frac{1}{a^3M_p^2}{\left(\frac{a^3} {H}\dot{\chi}\dot{\sigma}\right)}^.\big]Q_{\sigma}=0\,, \label{diffequchi} \\ \nonumber
\ddot{Q_{\sigma}}+3H\dot{Q_{\sigma}}+\frac{k^2}{a^2}Q_{\sigma}+\big[V_{,\sigma \sigma}-\frac{1}{a^3M_p^2}{\left(\frac{a^3}{H}\dot{\sigma}^2\right)}^.\big]Q_{\sigma}
&+\big[V_{,\sigma \phi}-\frac{1}{a^3M_p^2}{\left(\frac{a^3}{H}\dot{\sigma}\dot{\phi}\right)}^.\big]Q_{\phi}\nonumber\\
&+\big[V_{, \sigma \chi}-\frac{1}{a^3M_p^2}{\left(\frac{a^3} {H}\dot{\sigma}\dot{\chi}\right)}^.\big]Q_{\chi}=0\,, \label{diffequchi}
\end{align}
where $k=\frac{2\pi a}{\lambda}$ is the comoving wave number of the mode with physical wavelength $\lambda$,  ${()}^.\equiv\frac{d}{dt}()$, and $V_{,a}\equiv\frac{\partial V}{\partial a}, V_{,ab}\equiv\frac{\partial^2 V}{\partial a \partial b}$. We define
\begin{equation}\label{CIJ}
    C_{IJ}\equiv V_{,IJ}-\frac{1}{a^3M_p^2}{\left(\frac{a^3}{H}\dot{I}\dot{J}\right)}^. ,
\end{equation}
where \textit{I} and \textit{J} can be each of our fields. The set of equations can be then written as
\begin{align}
\ddot{Q}_{\phi}+3H\dot{Q}_{\phi}+\left(\frac{k^2}{a^2}+C_{\phi\phi}\right)Q_{\phi}+C_{\phi\chi}Q_{\chi}+C_{\phi\sigma}Q_{\sigma}&=0\,, \label{diffequphibr}\\
\ddot{Q}_{\chi}+3H\dot{Q}_{\phi}+\left(\frac{k^2}{a^2}+C_{\chi\chi}\right)Q_{\chi}+C_{\chi\phi}Q_{\phi}+C_{\chi\sigma}Q_{\sigma}&=0\,, \label{diffequchibr}\\
\ddot{Q}_{\sigma}+3H\dot{Q}_{\sigma}+\left(\frac{k^2}{a^2}+C_{\sigma\sigma}\right)Q_{\sigma}+C_{\sigma\phi}Q_{\phi}+C_{\sigma\chi}Q_{\chi}&=0\,, \label{diffequsigmabr}
\end{align}
where
\begin{align}
C_{\phi\phi}&= V_{,\phi\phi}+\frac{3\dot{\phi}^2}{M_p^2}+\frac{2\dot{\phi}V_{\phi}}{M_p^2H}-\frac{\dot{\phi}^2(\dot{\phi}^2+\dot{\chi}^2+\dot{\sigma}^2)}{2M_p^4H^2}\,,\label{Cphiphi}\\
C_{\chi\chi}&= V_{,\chi\chi} +\frac{3\dot{\chi}^2}{M_p^2}+\frac{2\dot{\chi}V_{\chi}}{M_p^2H}-\frac{\dot{\chi}^2(\dot{\phi}^2+\dot{\chi}^2+\dot{\sigma}^2)}{2M_p^4H^2}\,, \label{Cchichi}\\
C_{\sigma\sigma}&= V_{,\sigma\sigma}+\frac{3\dot{\sigma}^2}{M_p^2}+\frac{2\dot{\sigma}V_{\sigma}}{M_p^2H}-\frac{\dot{\sigma}^2(\dot{\phi}^2+\dot{\chi}^2+\dot{\sigma}^2)}{2M_p^4H^2}\,, \label{Csigmasigma}\\
C_{\phi\chi}&=V_{,\phi\chi}+ \frac{3\dot{\phi}\dot{\chi}}{M_p^2}+\frac{\dot{\phi}V_{\chi}+\dot{\chi}V_{\phi}}{M_p^2H}-\frac{\dot{\phi}\dot{\chi}(\dot{\phi}^2+\dot{\chi}^2+\dot{\sigma}^2)}{2M_p^4H^2}= C_{\chi\phi}\,,\label{Cphichi}\\
C_{\phi\sigma}& = V_{,\phi\sigma}+\frac{3\dot{\phi}\dot{\sigma}}{M_p^2}+\frac{\dot{\phi}V_{\sigma}+\dot{\sigma}V_{\phi}}{M_p^2H}-\frac{\dot{\phi}\dot{\sigma}(\dot{\phi}^2+\dot{\chi}^2+\dot{\sigma}^2)}{2M_p^4H^2}= C_{\sigma\phi}\,, \label{Cphisigma}\\
C_{\chi\sigma}&  =V_{,\chi\sigma}+ \frac{3\dot{\chi}\dot{\sigma}}{M_p^2}+\frac{\dot{\chi}V_{\sigma}+\dot{\sigma}V_{\chi}}{M_p^2H}-\frac{\dot{\chi}\dot{\sigma}(\dot{\phi}^2+\dot{\chi}^2+\dot{\sigma}^2)}{2M_p^4H^2}= C_{\sigma\chi} \,.\label{Cchisigma}\\ \nonumber
\end{align}
\subsection{Curvature and isocurvature directions}
\begin{figure}
     \centering     \includegraphics[width=\textwidth]{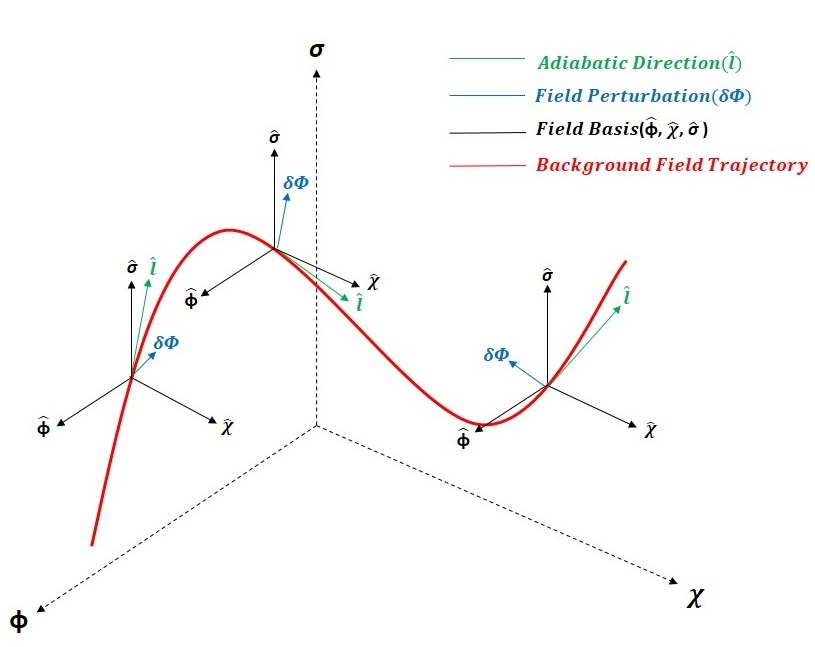}
\caption{This figure shows the adiabatic direction (green vectors), defined at each point as the tangent direction to the field trajectory (shown in red). At each point on the field trajectory we have a field basis (shown in black), and at each point we have a field perturbation vector (shown in blue vectors). We should transform this vector to a new basis at each point of which one of its components is along the adiabatic direction. The other two components should be on the surface orthogonal to $\hat{l}$.}
    \label{traj-ins-basis}
 \end{figure}

To capture curvature perturbations in multifield inflation models—which later manifest as temperature fluctuations on the CMB—one must project the perturbations along the tangent direction of the field trajectory\cite{Gordon:2000hv}(see Fig. \ref{traj-ins-basis}). This projection is typically achieved in the literature through a coordinate transformation in field space, from the field basis to the so-called kinematic basis. As the name suggests, in this basis, perturbations are decomposed along the directions of the velocity vector, the acceleration vector, and higher-order kinematic quantities. In two-field models, only two directions are relevant: the tangent to the field trajectory (aligned with the velocity vector) and the perpendicular direction (aligned with the acceleration vector). The tangent direction corresponds to the adiabatic perturbation, while the perpendicular direction captures the sole isocurvature mode.

\begin{figure}
     \centering     \includegraphics[width=\textwidth]{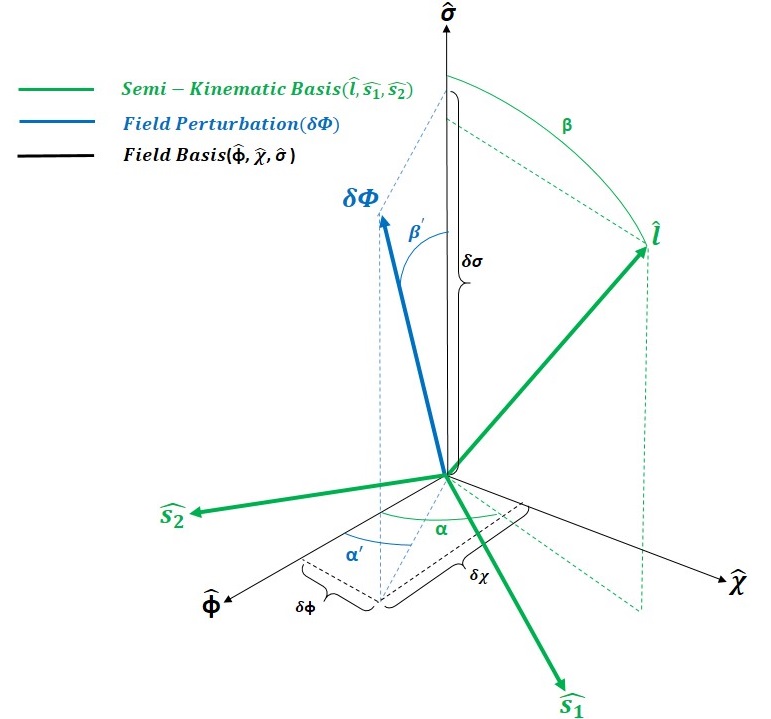}
     \caption{Field basis (black) and semikinematic basis (green) are shown. The adiabatic direction is determined by two angles, $\alpha$ and $\beta$. The field perturbation vector is also shown in blue. This vector's direction is also determined by two angles, $\alpha'$ and $\beta'$.}
     \label{traj-andgles}
 \end{figure}

However, for models with $N > 2$ fields, there is an additional feature: the freedom to choose isocurvature directions in field space. The only constraint is that the $N-1$ isocurvature perturbations must remain orthogonal to the tangent vector, which corresponds to the adiabatic mode. Considering this point, the question arises that in more-than-two-field models, in which the kinematic basis is not the only choice for isocurvature modes, what basis should we choose, and what are the advantages and disadvantages of each choice? Here we elaborate on this issue:
\begin{itemize}
    \item In the kinematic basis, one isocurvature perturbation is aligned with the acceleration direction, and this is the only isocurvature mode that couples to the adiabatic perturbation. The remaining  $N-2$ isocurvature modes remain decoupled from the adiabatic perturbations at each moment. Therefore, by choosing the kinematic basis, one can follow the coupling between the adiabatic and the isocurvature mode. However, if the two-dimensional plane that is spanned by the adiabatic direction and its acceleration changes, then different isocurvature modes will be coupled to the adiabatic mode in turn.
\item Depending on the background field trajectory, the acceleration of the field vector may vanish at some times, or even at all times (e.g., if the model is a single-field model at the background level, and the other two modes are turned on at the perturbative level). In such circumstances, the kinematic basis is ill-defined, since we basically cannot define the acceleration direction.
\item Finding the acceleration direction, and the second isocurvature direction, is not a practically easy task. The transformation matrix from the field basis to the kinematic basis in the three-field case is not a simple matrix, and this makes the calculations complicated. We have obtained this matrix in Appendix \ref{Appendix-A}.
\item The freedom to choose isocurvature directions in the hypersurface orthogonal to the adiabatic direction implies that the isocurvature perturbations computed during inflation are not, by themselves, observables at late time. Directly observable isocurvature modes are formed later, after reheating, when these initial perturbations decay into the entropy perturbations of the radiation-dominated phase. In this process, the transfer matrix adapts to the choice of initial isocurvature basis, thereby ensuring the invariance of the physical correlation functions in the radiation era (see Appendix \ref{Appendix-A}). We investigate the relation to observables more thoroughly in Sec. \ref{RTO}.
\end{itemize}
According to the above comments, we choose a basis that we call the semikinematic basis, which has the adiabatic component parallel to the tangent to the trajectory, but the two other directions are not along the higher-order kinematic quantities. The other two directions in this basis would be obtained from rotating the other two field directions with the same rotation matrix that cast the first field direction along the tangent to the trajectory. Field perturbations in this basis are given by the transformation matrix $\mathcal{M}_s$, defined as
\begin{align}
\begin{pmatrix}
       \delta l \\
       \delta s_1\\
       \delta s_2
\end{pmatrix}
       &=\mathcal{M}_s
\begin{pmatrix}
     \delta \phi \\
       \delta \chi \\
       \delta \sigma
\end{pmatrix},
\label{MSvstrans}
\end{align}
\begin{align}
\mathcal{M}_s=
     \begin{pmatrix}
       \sin\beta\cos\alpha & \sin\beta\sin\alpha & \cos\beta\\
       -\sin\alpha & \cos\alpha & 0\\
        -\cos\beta\cos\alpha & -\cos\beta\sin\alpha & \sin\beta
     \end{pmatrix},
     \label{Mdef}
\end{align}
where $\beta$ and $\alpha$ are the angles determining the adiabatic direction at each point, shown in Fig. \ref{traj-andgles},
\begin{align}
    \cos\beta &\equiv \frac{\dot{\sigma}}{\dot{l}}, \label{cosbeta}\\
    \sin\beta &\equiv \frac{\sqrt{\dot{\phi}^2+\dot{\chi}^2}}{\dot{l}}, \label{sinbeta}\\
    \cos\alpha &\equiv \frac{\dot{\phi}}{\sqrt{\dot{\phi}^2+\dot{\chi}^2}}, \label{cosalpha}\\
    \sin\alpha &\equiv \frac{\dot{\chi}}{\sqrt{\dot{\phi}^2+\dot{\chi}^2}}. \label{sinalpha}
\end{align}
In Appendix \ref{Appendix-A} the derivation of $\mathcal{M}_s$ is shown. The adiabatic unit vector, $\hat{l}$, tangent to the background trajectory, is defined as
\begin{equation}
    \hat{l}^a = \frac{\dot{\phi}^a}{\dot{l}}, \hspace{0.7cm} \dot{l} = \dot{\phi}^2 + \dot{\chi}^2 + \dot{\sigma}^2,
\end{equation}
where $a = 1,2,3$, which correspond to $\phi$, $\chi$, and $\sigma$. The isocurvature directions, perpendicular to $\hat{l}^a$, are $\hat{s}_1$ and $\hat{s}_2$.
We can project the equation of motion of the fields, Eqs. \eqref{backequphi}—\eqref{backequsig}, in the direction of this basis,
\begin{align}
\ddot{l} + 3H\dot{l} + V_{,l} &= 0, \label{BEql} \\
\hat{s}_{1_a} \ddot{\phi^a}  + V_{,s_1} &= 0, \\
\hat{s}_{2a} \ddot{\phi^a} + V_{,s_2} &= 0,
\end{align}
where  $V_{,l} = \hat{l}^a V_{,a}$, $V_{,s_1} = \hat{s}_1^a V_{,a}$, and $V_{,s_2} = \hat{s}_2^a V_{,a}$.
We also define a useful quantity, which is usually called the covariant field acceleration in the literature,
\begin{equation}
    \eta^a = \frac{\ddot{\phi}^a}{H \dot{l}}.
\end{equation}
It can be decomposed to kinematic basis components,
\begin{align}
    \hat{l}_a\eta^a &= \eta_{||} = \frac{\dot{\phi}_a\ddot{\phi}^a}{H \dot{l}^2} = \frac{\ddot{l}}{H \dot{l}} \label{seconsSRP}\\
    \hat{s}_{1 a}\eta^a &= \eta_{\perp 1}=  \frac{s_{1 a}\ddot{\phi}^a}{H \dot{l}} = -\frac{V_{,s_1}}{H \dot{l}} \label{TurnRate1}\\
     \hat{s}_{2 a}\eta^a &= \eta_{\perp 2}= \frac{s_{2 a}\ddot{\phi}^a}{H \dot{l}} = -\frac{V_{,s_2}}{H \dot{l}} \label{TurnRate2}
\end{align}
As is obvious, $\eta_{\perp 1}$ and $\eta_{\perp 2}$ characterize the turning rate of the field trajectory in comparison with the Hubble rate. In this sense, when these quantities are larger than unity, the trajectory is undergoing a rapid turn.

To derive the perturbation equations in this basis, we should have a rotation matrix, $\mathcal{M}_s$, by which at each point on the field trajectory we can transform ($\hat{\phi}, \hat{\chi}, \hat{\sigma}$) to ($\hat{l}, \hat{s_1}, \hat{s_2}$), and also transform Mukhanov-Sasaki variables $Q_{\phi}, Q_{\chi},Q_{\sigma}$ to $Q_l, \delta s_1, \delta s_2$,
\begin{align}
\begin{pmatrix}
       Q_l \\
       \delta s_1\\
       \delta s_2
\end{pmatrix}
       &=\mathcal{M}_s
\begin{pmatrix}
       Q_{\phi} \\
       Q_{\chi} \\
       Q_{\sigma}
\end{pmatrix} \label{MSvstrans1}.
\end{align}
$Q_l$ is defined as
\begin{equation}\label{Qldef}
    Q_l = \delta l+\frac{\dot{l}}{H}\Phi\,,
\end{equation}
and, in the comoving gauge, it is directly related to the three-dimensional curvature of the constant time spacelike slices, $\mathcal{R}$,
\begin{equation}\label{Rdef}
    \mathcal{R} \equiv \frac{H}{\dot{l}}Q_l.
\end{equation}
,$\delta s_1$ and $\delta s_2$ are called the isocurvature perturbations and are gauge invariant. Analogous to $\mathcal{R}$, we can define entropy perturbations from the isocurvature perturbations as
\begin{align}
    \mathcal{S}_1 \equiv \frac{H}{\dot{l}}\delta s_l, \label{isocp1def}\\
    \mathcal{S}_2 \equiv \frac{H}{\dot{l}}\delta s_2 \label{isocp2def}.
\end{align}
Using the rotation matrix, $\mathcal{M}$, we can now write Eqs. \eqref{diffequphibr} — \eqref{diffequsigmabr} in terms of the new variables,\footnote{We can reduce these three-field perturbative equations to the two field case, if we put for example $\sigma$ in its minimum, and therefore $\beta = \frac{\pi}{2}$, and also set the derivatives of the potential w.r.t. $s_2$ vanish. It can be shown that these reduced equations are the same as the Eqs. (34)—(35) in \cite{Lalak:2007vi} in the flat field space limit.}
\begin{align}
     \ddot{Q_l} &+ 3H\dot{Q_l} + \left(\frac{k^2}{a^2}+C_{ll}\right)Q_l + 2\frac{V_{,s_1}}{\dot{l}}\dot{\delta s_1} + 2\frac{V_{,s_2}}{\dot{l}}\dot{\delta s_2} + C_{ls_1}\delta s_1 + C_{ls_2}\delta s_2 = 0\,, \label{diffequQl}\\
    \ddot{\delta s_1} &+ 3H\dot{\delta s_1} + \left(\frac{k^2}{a^2}+C_{s_1s_1}\right)\delta s_1 - 2\frac{V_{,s_1}}{\dot{l}}\dot{Q_l} + 2\frac{V_{,s_1}\cot \beta}{\dot{l}}  \dot{\delta s_2} +  C_{s_1l} Q_l  + C_{s_1s_2} \delta s_2= 0\,, \label{diffequS1}\\
       \ddot{\delta s_2} &+ 3H\dot{\delta s_2}+  \left(\frac{k^2}{a^2}+C_{s_2s_2}\right)\delta s_2 - 2\frac{V_{,s_2}}{\dot{l}} \dot{Q_l} - 2\frac{V_{,s_1}\cot \beta}{\dot{l}}  \dot{\delta s_1}  +  C_{s_2l} Q_l  + C_{s_2s_1} \delta s_1= 0, \label{diffequS2}
\end{align}
where
\begin{align}
    C_{ll} &= V_{,ll} -\frac{\dot{l}^4}{2 H^2}+ 3\dot{l}^2 +\frac{2 V_{,l}\dot{l}}{{H}} -\frac{V_{,s1}^2+V_{,s2}^2}{\dot{l}^2}\,, \label{Cll} \\
    C_{s_1 s_1} &= V_{,s_1 s_1}-\frac{V_{,s_1}^2 \csc ^2\beta }{\dot{l}^2}\,, \label{Cs1s1} \\
    C_{s_2 s_2} &= V_{,s_2 s_2}-\frac{V_{,s_1}^2 \cot ^2 \beta +V_{,s_2}^2}{\dot{l}^2}\,, \label{Cs2s2} \\
      C_{l s_1} &= 2V_{,ls_1} +\frac{V_{,s_1}\dot{l}}{H}+\frac{6HV_{,s_1}}{\dot{l}}+\frac{2 V_{,l} V_{,s_1}-2 V_{,s_1} V_{,s_2} \cot \beta}{\dot{l}^2}\,,\label{Cls1} \\
          C_{l s_2} &= 2V_{,ls_2} +\frac{V_{,s_2}\dot{l}}{H} +\frac{6 H V_{,s_2}}{\dot{l}} +\frac{2 V_l V_{,s_2}+(V_{,s_1}^2-V_{,s_2}^2) \cot\beta}{\dot{l}^2}\,, \label{Cls2}   \\
    C_{s_1 l} &=\frac{V_{,s_1}\dot{l}}{H}-\frac{6 H V_{,s_1}}{\dot{l}}-\frac{2 V_l V_{,s_1} }{\dot{l}^2}\,, \label{Cs1l}    \\
    C_{s_2 l} &= -\cot\beta V_{,ll}+\frac{V_{,s_2}\dot{l}}{H}- \frac{6 H V_{,s_2}}{\dot{l}} - \frac{2 V_l V_{,s_2}-(V_{,s_1}^2+V_{,s_2}^2) \cot\beta}{\dot{l}^2}\,, \label{Cs2l}   \\
    C_{s_1 s_2} &= V_{,s_1s_2}+ \cot \beta V_{,s_1 l}+\frac{6 H V_{,s_1}  \cot\beta}{\dot{l}} +\frac{2V_{,s_1}(V_l\cot \beta -V_{,s_2}\csc^2 \beta )}{\dot{l}^2}\,, \label{Cs1s2}    \\
    \label{Cs2s1}
    C_{s_2 s_1} &= V_{,s_2s_1}-\cot \beta V_{,s_1 l}-\frac{6 H V_{,s_1}  \cot\beta}{\dot{l}} -\frac{2V_{,s_1}(V_l\cot \beta -V_{,s_2}\cot^2 \beta )}{\dot{l}^2}\,.
\end{align}

See Appendix \ref{Appendix-B} for the more detailed derivation of \eqref{diffequQl}—\eqref{diffequS2}. As it can be seen, these equations are in the most general form, by which we mean all of the curvature and isocurvature modes are coupled together. Couplings of different modes, in general, depend on the derivatives of potential and the quantities defined in \eqref{Cll} — \eqref{Cs2s1}, all of which depend on the background trajectory.

\subsection{Observables} \label{subsec:observables}
We are interested in the power spectra of the curvature and isocurvature modes, $ \mathcal{P}_{\mathcal{R}}$, $\mathcal{P}_{\mathcal{S}_1}$, and $\mathcal{P}_{\mathcal{S}_2}$. We solve Eqs. \eqref{diffequQl}—\eqref{diffequS2} numerically around the background trajectory. Then, using \eqref{Rdef}—\eqref{isocp2def} the power spectra can be obtained from
\begin{align}
    \mathcal{P}_{\mathcal{R}} = \frac{k_i^3}{2\pi^2}|\mathcal{R}|^2\,, \label{defPRN}\\
    \mathcal{P}_{\mathcal{S}_1} = \frac{k_i^3}{2\pi^2}|\mathcal{S}_1|^2\,, \label{defPS1N}\\
    \mathcal{P}_{\mathcal{S}_2} = \frac{k_i^3}{2\pi^2}|\mathcal{S}_2|^2\,, \label{defPS2N}
\end{align}
where $k_i$ is a specific $k$ that we are interested in, which in our case is just the modes that exit the horizon after 60 e-folds before the end of inflation. The ones that exit the horizon between 60 to 50 e-folds before the end of inflation correspond to the CMB scales.
Another quantity that would be useful is the correlation between these different modes, defined as
\begin{align}
    \mathcal{C}_{\mathcal{R} \mathcal{S}_1} = \frac{k_i^3}{2\pi^2} \mathcal{R}\mathcal{S}_1^{\dagger}\,, \label{defCRS1N}\\
    \mathcal{C}_{\mathcal{R} \mathcal{S}_2} = \frac{k_i^3}{2\pi^2}\mathcal{R}\mathcal{S}_2^{\dagger}\,, \label{defCRS2N}\\
    \mathcal{C}_{ \mathcal{S}_1 \mathcal{S}_2} = \frac{k_i^3}{2\pi^2}\mathcal{S}_1 \mathcal{S}_2^{\dagger} \label{defCS1S2N}.
\end{align}
An important point that should be considered for solving the Eqs. \eqref{diffequQl}—\eqref{diffequS2} is the way we set the initial conditions. The first approach that comes to mind is assigning the Bunch-Davis (BD) initial condition to all curvature and isocurvature modes simultaneously and solving the equations numerically. However, to take into account the statistical independence of the adiabatic and the isocurvature perturbations deep
inside the Hubble radius, the correct way to proceed is to solve the equations three times, each time setting only one of the modes in the BD initial condition, setting the rest equal to zero. For each initial condition we calculate the contribution to the two-point correlation function then add them up to obtain the total two-point correlation functions.

The relative correlation coefficients, then are defined as
\begin{equation}
\tilde{\mathcal{C}}_{ij}=\frac{|\mathcal{C}_{ij}|}{\sqrt{\mathcal{P}_{i}\mathcal{P}_{j}}}\,,
\label{Cij}
\end{equation}
where $i,j\in\{\mathcal{R},\mathcal{S}_1,\mathcal{S}_2\}$.
The value of these quantities is between zero and one, and it indicates to what extent the final curvature perturbations result from the interactions with the isocurvature perturbations \cite{Lalak:2007vi}. \\


\section{Results}\label{sec4}

In this section, we will solve the background and linear perturbative level equations for three different cases. The first case is inspired by the M-flation model \cite{Ashoorioon:2009wa,Ashoorioon:2009sr}, and is, in fact, a two-field case. By investigating this case we intend to check the consistency of the three-field formulation in reducing to the known two-field formalism. The second and third cases that we investigate numerically are, on the other hand, the cases in which all three DoFs are dynamical during inflation. In the second case, different fields do not directly interact with each other, while in the third case direct interaction between fields is also included. \\

\subsection{Initial conditions}
   Equations of motion for the fields are second order, and each of them requires two boundary conditions. We are going to set the initial conditions at 80 e-folds before the end of inflation, $N_e = N_{e_i} = -80$, thus we can use the slow-roll approximation of Eqs. \eqref{backequphi}—\eqref{backequH} to set the initial conditions for derivatives of the fields
\begin{align}
\phi(N_{e_i}) &=\phi_i\,, \\
\chi(N_{e_i}) &=\chi_i\,, \\
\sigma(N_{e_i}) &=\sigma_i\,,\\
   \phi'(N_{e_i}) &=  \frac{V_{, \phi}(\phi_i,\chi_i,\sigma_i)}{3V(\phi_i,\chi_i,\sigma_i)}\,, \label{backequphiNSR} \\
    \chi'(N_{e_i}) &=  \frac{V_{, \chi}(\phi_i,\chi_i,\sigma_i)}{3V(\phi_i,\chi_i,\sigma_i)}\,, \label{backequchiNSR} \\
         \sigma'(N_{e_i}) &=  \frac{V_{, \sigma}(\phi_i,\chi_i,\sigma_i)}{3V(\phi_i,\chi_i,\sigma_i)}\,,  \label{backequsigNSR}
\end{align}
where we used the slow-roll approximation of \eqref{backequphiN}—\eqref{backequsigN}. The initial condition for $H$ is determined from \eqref{backequHN},
\begin{equation}
    H(N_{e_i}) = \sqrt{-\frac{2V(\phi_i , \chi_i , \sigma_i)}{\phi^{'2}_i+\chi^{'2}_i+\sigma^{'2}_i-6}}\,.
\end{equation}
After obtaining the background trajectory of the field space, solving Eqs. \eqref{diffequQl}—\eqref{diffequS2}, and using \eqref{defPRN}—\eqref{defCS1S2N}, we can obtain the power spectra of curvature and isocurvature modes, and also the relative correlation function. \\

In the $a \rightarrow 0$ limit, which corresponds to $N_e \rightarrow -\infty$, $\frac{k^2}{a^2}$ will be the dominant term in  \eqref{diffequQl}-\eqref{diffequS2}, and we can neglect other terms. This means that in that limit $Q_l$, $\delta s_1$ and $\delta s_2$ are not coupled with each other. Moreover, since orthogonal transformations in the field space preserve the canonical kinetic term for Mukhanov-Sasaki variables, at that limit equations will become
\begin{align}
    Q''_l+\big(k^2 -\frac{a''}{a} \big)Q_l = 0, \\
    \delta s_1''+\big(k^2 -\frac{a''}{a} \big)\delta s_1 = 0, \\
    \delta s_2''+\big(k^2 -\frac{a''}{a} \big)\delta s_2 = 0,
\end{align}
where prime denotes differentiation with respect to the conformal time. These equations simply lead to the Bunch-Davis vacuum that should be normalized by $\frac{1}{\sqrt{2k}}$ to account for the Wronskian condition. Therefore, by assuming that this argument holds with sufficiently good precision for $N_e=-80$, we are allowed to choose the Bunch-Davis initial condition. We verified this assumption by explicitly checking it out using the background value of the coefficients in the perturbative equations.

As already explained in Sec. \ref{subsec:observables}, we solve the linear perturbation equations three times, each time setting only one of the modes in the BD initial condition, setting the rest equal to zero. For each initial condition we calculate the contribution to the two-point correlation function then add them up to obtain the total two-point correlation functions.
\begin{enumerate}
    \item
    \begin{align*}
        Q_l(N_{e_i}) &= \frac{1}{a(N_{e_i})\sqrt{2k}}\,, \hspace{1cm}  Q_l'(N_{e_i}) = -\frac{ik + H(N_{e_i})a'(N_{e_i})}{ H(N_{e_i})a(N_{e_i})^2\sqrt{2k}} \\ \nonumber
        \delta s_1(N_{e_i}) &= 0\,, \hspace{2.5cm}   \delta s_1'(N_{e_i}) = 0 \\ \nonumber
        \delta s_2(N_{e_i}) &= 0\,, \hspace{2.5cm}   \delta s_2'(N_{e_i}) = 0
    \end{align*}
     \item
    \begin{align*}
        Q_l(N_{e_i}) &= 0\,, \hspace{2.5cm}  Q_l'(N_{e_i}) = 0 \\ \nonumber
        \delta s_1(N_{e_i}) &= \frac{1}{a(N_{e_i})\sqrt{2k}}\,, \hspace{1cm}   \delta s'_1(N_{e_i}) = -\frac{ik + H(N_{e_i})a'(N_{e_i})}{ H(N_{e_i})a(N_{e_i})^2\sqrt{2k}} \\ \nonumber
        \delta s_2(N_{e_i}) &= 0\,, \hspace{2.5cm}   \delta s_2'(N_{e_i}) = 0
    \end{align*}
     \item
    \begin{align*}
        Q_l(N_{e_i}) &= 0\,, \hspace{2.5cm}  Q_l'(N_{e_i}) = 0 \\ \nonumber
        \delta s_1(N_{e_i}) &= 0\,, \hspace{2.5cm}   \delta s'_1(N_{e_i}) = 0 \\ \nonumber
        \delta s_2(N_{e_i}) &= \frac{1}{a(N_{e_i})\sqrt{2k}}\,, \hspace{1cm}   \delta s_2'(N_{e_i}) = -\frac{ik + H(N_{e_i})a'(N_{e_i})}{ H(N_{e_i})a(N_{e_i})^2\sqrt{2k}}
    \end{align*}
\end{enumerate}

\subsection{Two-field case}
In this case, by putting the third field in the minimum of its potential manually, we have a two-field model at the background level. This case is motivated by the two-block case of the matrix inflation model \cite{Ashoorioon:2009wa,Ashoorioon:2009sr,Ashoorioon:2011ki,Ashoorioon:2014jja}. By investigating this case, we can also check if the three-field formulation correctly reduces to the two-field one. Different fields do not directly interact with each other and,  therefore the potential consists of the potential of each field,
\begin{equation}\label{potform}
    V(\phi, \chi, \sigma) =  V_{\phi}(\phi) +  V_{\chi}(\chi) + V_{\sigma}(\sigma).
\end{equation}
The potentials are chosen from one of the relevant cases in thes M-flation landscape \cite{Ashoorioon:2009wa},
\begin{align}
    V_{\phi}(\phi)&=\frac{\lambda_{\phi}}{4} \phi^2(\phi^2-\mu_{\phi})^2\,, \label{potentialphic2} \\
    V_{\chi}(\chi) &=\frac{\lambda_{\chi}}{4} \chi^2(\chi^2-\mu_{\chi})^2\,, \label{potentialchic2} \\
    V_{\sigma} (\sigma) &= \frac{1}{2}m \hspace{0.1cm} \sigma^2\,. \label{potentialsigc1} \\ \nonumber
\end{align}
In \cite{Ashoorioon:2009sr}, $\sigma$ represents one of the uncorrelated isocurvature modes.

\subsubsection{Background solution}
\begin{figure}
     \centering     \includegraphics[width=\textwidth]{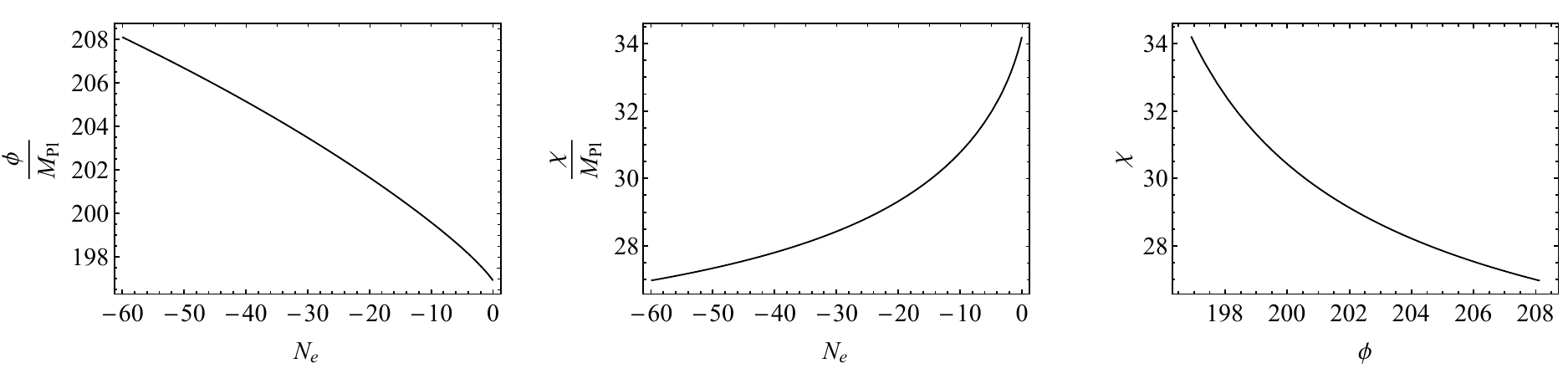}
\caption{Two-Field Case: the evolution of  $\phi$ and $\chi$ with respect to the number of e-folds, and the trajectory in the field space is shown. $N_e=0$ is the end of inflation.}
     \label{background-N-c1}
 \end{figure}

\begin{figure}
     \centering     \includegraphics[width=\textwidth]{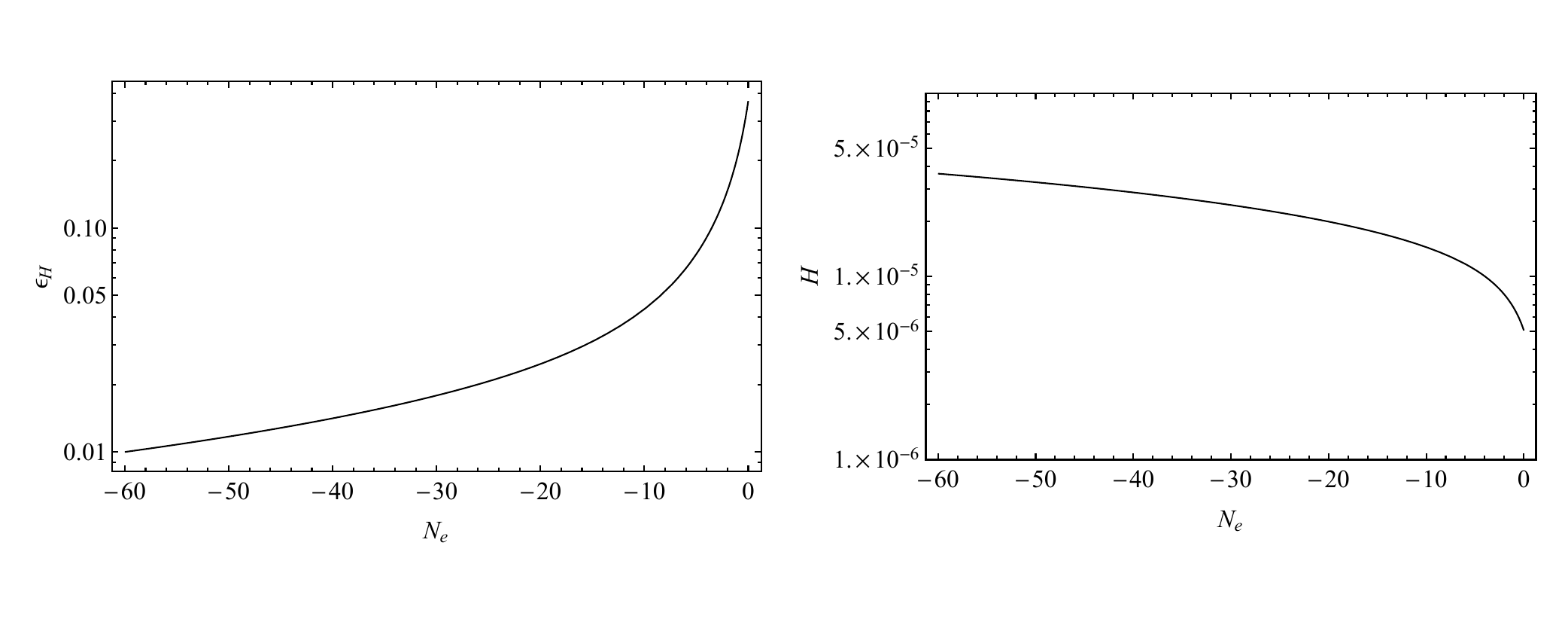}
\caption{Two-Field Case: the evolution of the Hubble parameter, \textit{H}, and the slow-roll parameter, $\epsilon_H$, with respect to the number of e-folds is shown.}
\label{HansSRPVSNe-c1}
 \end{figure}

Parameters and initial conditions are directly set from \cite{Ashoorioon:2009sr} and are shown in Table \ref{TableParas-1}.
The results for this case are shown in Figs. \ref{background-N-c1} and \ref{HansSRPVSNe-c1}. The left and middle plots of Fig. \ref{background-N-c1} show the evolution of fields $\phi$ and $\chi$ during inflation. On the other hand, $\sigma$ does not evolve during inflation, since it is placed at the minimum of its potential manually. Therefore, at the background level, this case is equivalent to a two-field model. The field trajectory is also shown in the right plot in Fig. \ref{background-N-c1}.  Figure \ref{HansSRPVSNe-c1} shows the evolution of the Hubble and slow-roll parameters.
\begin{table}[h!]
  \begin{center}
    \begin{tabular}{l|c|r} 
      \textbf{Parameters} & \textbf{Two-Field Case}\\
      \hline &\\
      $\lambda_{\phi}$ & $ 2.0 \times 10^{-15}$\\
      $\lambda_{\chi}$ & $\lambda_{\phi} \left(\frac{\mu_{\phi}}{\mu_{\chi}}\right)^2$  \\
      $\mu_{\phi}$ & $196.168 $\\
        $\mu_{\chi}$ & $36.000 $\\
        $\phi_i$ & $209.439 $\\
        $\chi_i$ & $26.678 $\\
         $m$ & $10^{-6} $\\
    \end{tabular}
     \caption{Two-Field Case: parameters of the potential and initial conditions. Since the third field is placed at the minimum of the potential, at the background level, its mass is irrelevant. $M_p$ is set to 1.}
    \label{TableParas-1}
  \end{center}
\end{table}

\subsubsection{Perturbation level}

\begin{figure}
     \centering     \includegraphics[width=\textwidth]{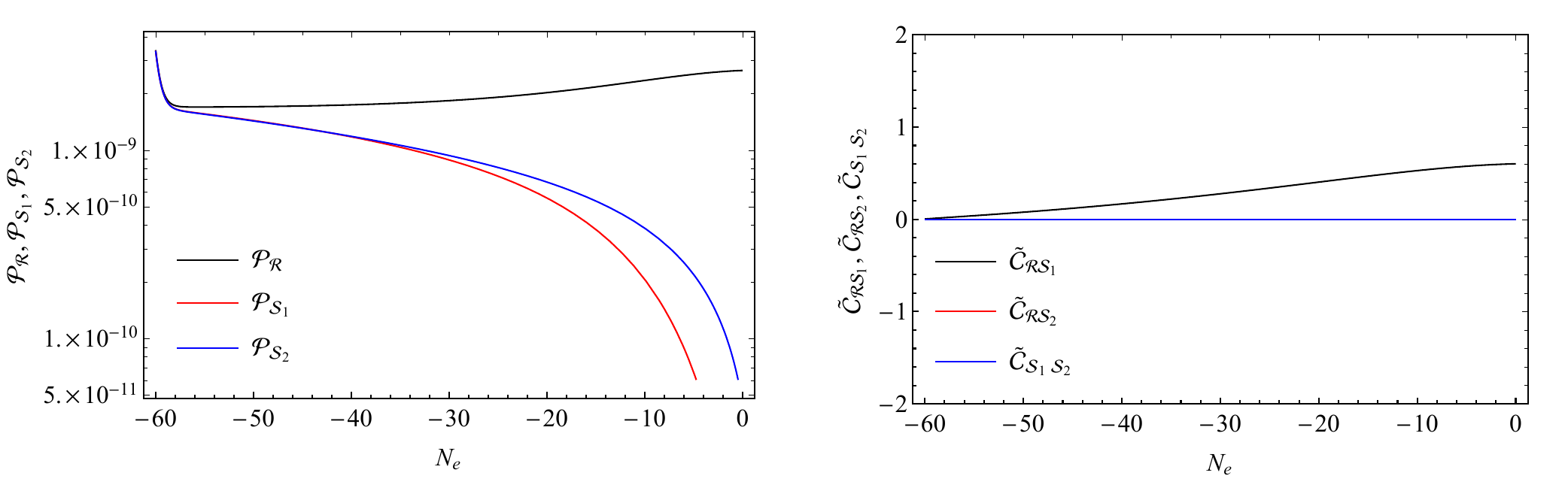}
     \caption{Two-Field Case: the evolution of the power spectra (left plot) of the curvature and isocurvature modes, and their correlations (right plot), defined in \eqref{Cij}, for the specific mode that exits the horizon 60 e-folds before the end of inflation, $ k= 0.002  \hspace{0.1cm} {\rm Mpc}^{-1}$, with respect to $N_e$.  The correlation including the second isocurvature mode vanishes ($\tilde{\mathcal{C}}_{\mathcal{R} \mathcal{S}_2}$ is invisible since it coincides with $\tilde{\mathcal{C}}_{\mathcal{S}_1 \mathcal{S}_2}$), and the correlation between the curvature and the first isocurvature mode increases during the inflation. The increase is related to the turn rate increase, shown in the left plot of Fig. \ref{PS-TR-ns-2}}
     \label{PRNe-1}
 \end{figure}

\begin{figure}
     \centering     \includegraphics[width=\textwidth]{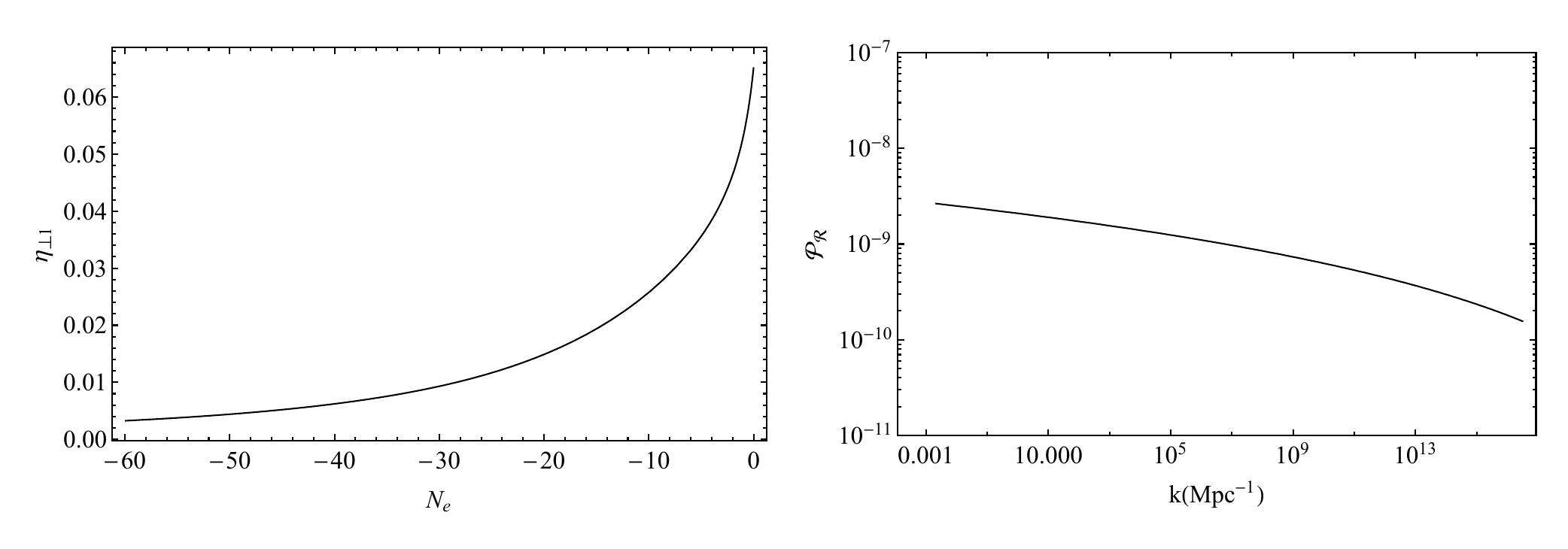}
\caption{Two-Field Case: t  he left plot shows the evolution of the turn rate of the trajectory during inflation. As is obvious, the trajectory does not experience rapid turns. The power spectrum of the curvature mode for different values of $k$, evaluated at the end of inflation, is shown in the right plot. The power spectra of the isocurvature modes have a similar shape with, respectively, 1 and 2 orders of magnitude lower amplitude.}
\label{PS-TR-ns-2}
 \end{figure}

We mentioned previously that the coefficients of different terms in the perturbation equations depend on the background trajectory. Since the background trajectory is in the $\sigma = 0$ surface, then we have $\beta = \frac{\pi}{2}$, and $\cos{\beta} = \cot{\beta} = 0$. Therefore, it can be seen that many of the coefficients in the equations vanish and the perturbation equations for this case take the forms
\begin{align}
        \ddot{Q_l} &+ 3H\dot{Q_l} + 2\frac{V_{,s_1}}{\dot{l}}\dot{\delta s_1} + \left(\frac{k^2}{a^2}+C_{ll}\right)Q_l + C_{ls_1}\delta s_1 = 0\,, \label{diffequQlC1}\\
    \ddot{\delta s_1} &+ 3H\dot{\delta s_1} - 2\frac{V_{,s_1}}{\dot{l}}\dot{Q_l} + \left(\frac{k^2}{a^2}+C_{s_1s_1}\right)\delta s_1 +  C_{s_1l} Q_l= 0\,, \label{diffequS1C1}\\
       \ddot{\delta s_2} &+ 3H\dot{\delta s_2}+  \left(\frac{k^2}{a^2}+C_{s_2s_2}\right)\delta s_2 = 0\,. \label{diffequS2C1}
\end{align}
It is obvious that the second isocurvature mode is completely decoupled from the other modes, as it was expected. In Fig. \ref{PRNe-1}, the evolution of the power spectra of the curvature and isocurvature modes (left plot), and the correlation between them (right plot), for a specific mode, which exits the horizon 60 e-folds before the end of inflation, $ k= 0.002  \hspace{0.1cm} {\rm Mpc}^{-1}$, with respect to $N_e$ are shown. As it could be easily observed, the amplitudes of the curvature perturbations grow even when the modes become superhorizon, which is an indication that the mode is fed by the correlated isocurvature perturbation. As it was expected, since the second isocurvature mode is perpendicular to the curvature and first isocurvature mode, the related correlation functions vanish. The correlation between the curvature and the first isocurvature mode increases during the inflation. This increase is related to the increasing turn rate, shown in the left plot of Fig. \ref{PS-TR-ns-2}.
The evolution of the power spectrum of the curvature mode, evaluated at the end of inflation, for different momentum modes is shown in the right plot. The value of the power spectrum is fixed on the known value $\mathcal{P}_{\mathcal{R}} \sim 2.1 \times 10^{-9}$ for the mode that exits at about 60 e-folds before the end of inflation. The power spectra of the isocurvature modes evolve similarly to the curvature mode with respect to $\textit{k}$, and their values for the mode exiting the horizon at the CMB scale are, respectively, $\mathcal{P}_{\mathcal{S}_1} \sim  10^{-10}$ and $\mathcal{P}_{\mathcal{S}_2} \sim 10^{-11}$. Amplitudes of the isocurvature modes are constrained by CMB. According to Planck data \cite{Planck:2018jri} $\beta_{iso}$, defined as $\frac{\mathcal{P}_{\mathcal{S}}}{\mathcal{P}_{\mathcal{S}_1}+\mathcal{P}_{\mathcal{R}}}$, can be at most at order of 0.1 for $\textit{k} = 0.002 \hspace{0.1cm} \rm Mpc^{-1}$. However, we should be careful that the constraints on isocurvature modes from CMB are for their value as the initial condition of the perturbations after inflation in the early radiation-dominated era. Therefore, the effect of these constraints on the amplitude of the isocurvature mode power spectrum at the end of inflation depends on the reheating model and the following dynamics. This issue will be investigated more in Sec. \ref{RTO}

\subsection{Three-field noninteracting case}

For the second case, we chose the potential in such a way that all three fields have dynamics during the inflation, so it is really a three-field model. Since the fields do not interact with each other, the potential form is just like \eqref{potform}, and we assume
\begin{align}
    V_{\phi}(\phi)&=\frac{\lambda_{\phi}}{4} \phi^2(\phi^2-\mu_{\phi})^2, \label{potentialphic2} \\
    V_{\chi}(\chi) &=\frac{\lambda_{\chi}}{4} \chi^2(\chi^2-\mu_{\chi})^2, \label{potentialchic2} \\
    V_{\sigma} (\sigma) &= \frac{\lambda_{\sigma}}{4} \sigma^2(\sigma^2-\mu_{\sigma})^2.
    \label{potentialsigc2} \\ \nonumber
\end{align}
\subsubsection{Background solution}
\begin{figure}
     \centering     \includegraphics[width=\textwidth]{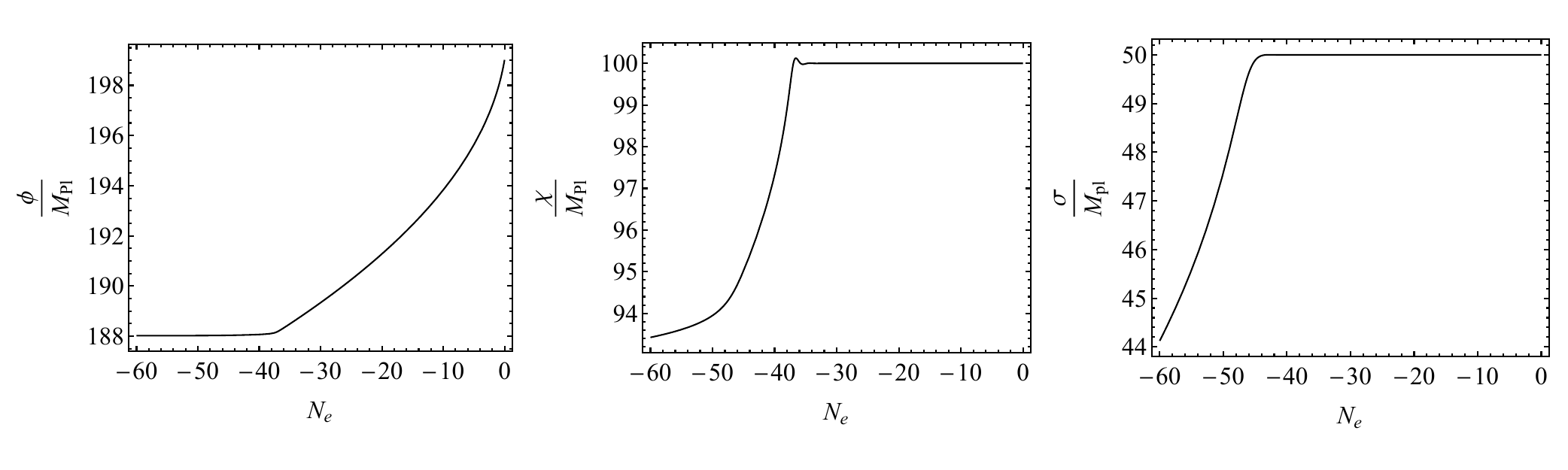}
\caption{Three-field noninteracting case: the evolution of the three scalar fields, $\phi$, $\chi$, and $\sigma$, is shown with respect to the number of e-folds. $N_e=0$ is the end of inflation}
     \label{FieldsVSNe-3RTST}
 \end{figure}

\begin{figure}
     \centering     \includegraphics[width=\textwidth]{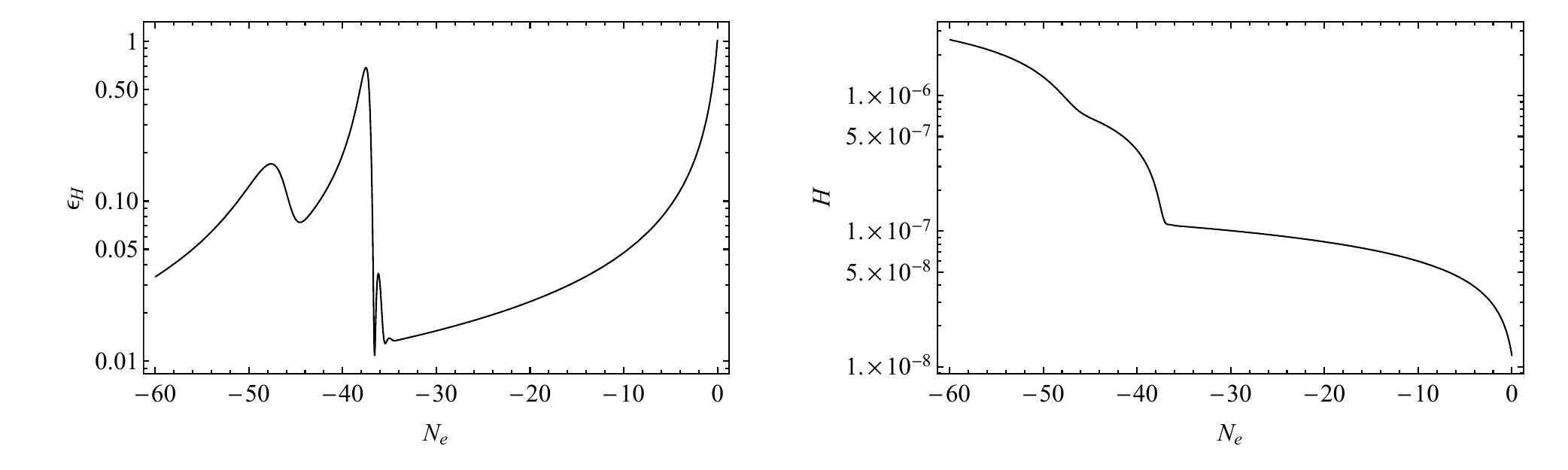}
\caption{Three-field noninteracting case I: the evolution of the Hubble parameter, \textit{H}, and the slow-roll parameter, $\epsilon_H$, is shown.}
\label{HepsVSNe-3RTS}
 \end{figure}

\begin{figure}
     \centering     \includegraphics[width=\textwidth]{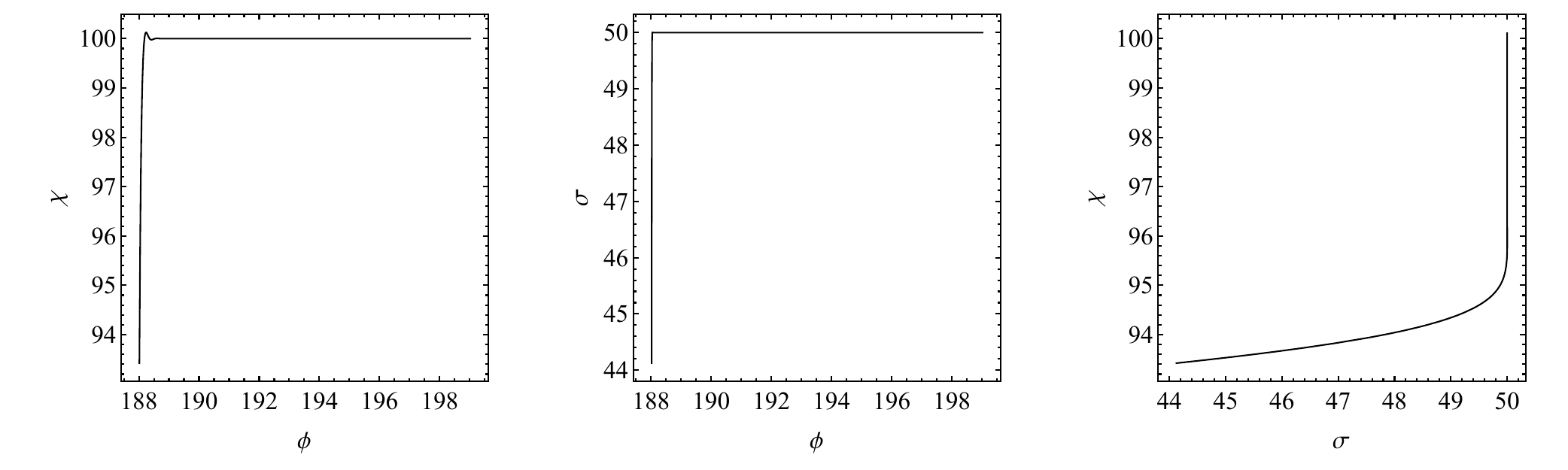}
\caption{Three-field noninteracting case I: 2D projections of the background trajectory in the field space}
     \label{FieldTraj2d-3RTST}
 \end{figure}

\begin{figure}
     \centering     \includegraphics[width=0.5\textwidth]{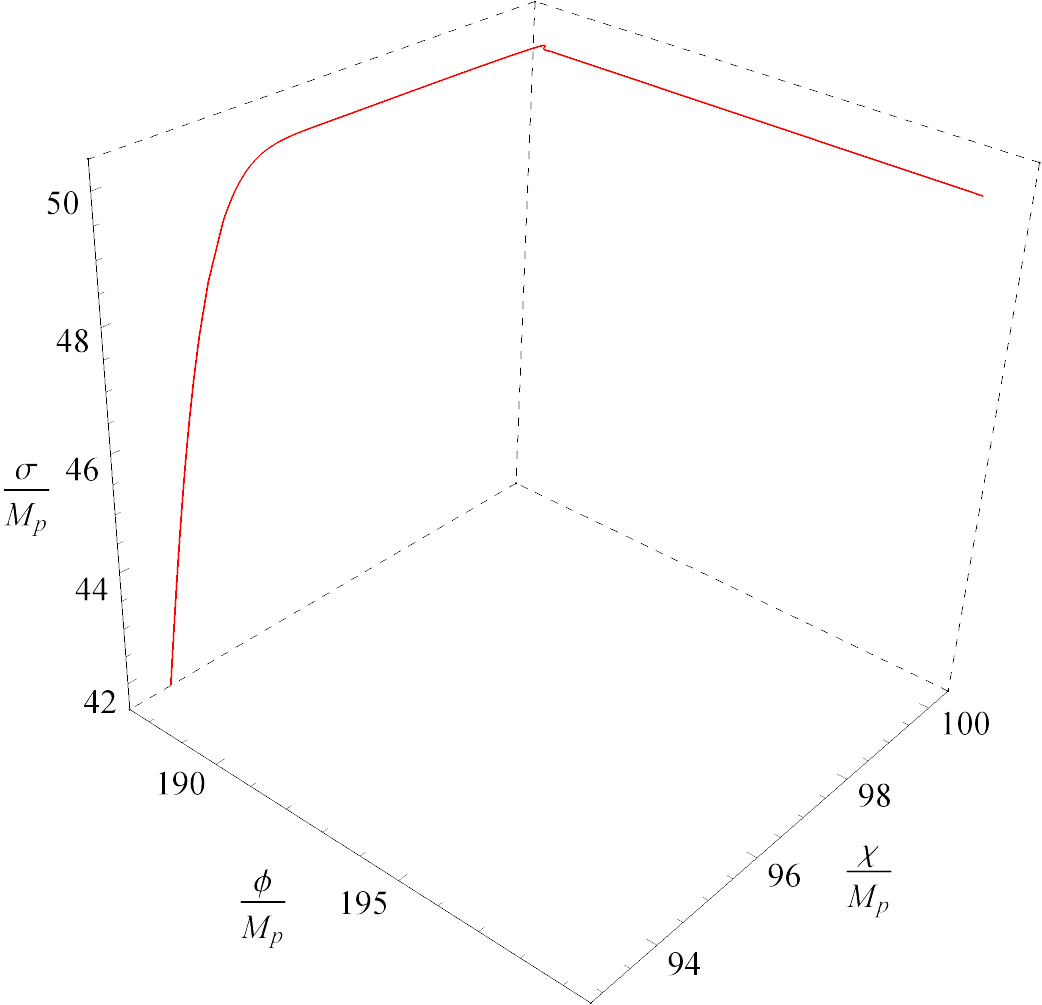}
\caption{Three-field noninteracting case I: the 3D background trajectory in the field space.}
     \label{FieldTraj3d-3RTST}
 \end{figure}

The parameters and initial conditions of this case are shown in Table  \ref{Paras-2}.
\begin{table}[h!]
  \begin{center}
    \begin{tabular}{l|c|r} 
      \textbf{Parameters} & \textbf{Three-Field Noninteracting Case I}\\
      \hline
      &\\
      $\lambda_{\phi}$ & $ 0.298 \times 10^{-19}$\\
      $\lambda_{\chi}$ & $200  \lambda_{\phi} \left(\frac{\mu_{\phi}}{\mu_{\chi}}\right)^2$  \\
       $\lambda_{\sigma}$ & $2200 \lambda_{\phi} \left(\frac{\mu_{\phi}}{\mu_{\sigma}}\right)^2$  \\
      $\mu_{\phi}$ & 200.0\\
        $\mu_{\chi}$ & 100.0\\
         $\mu_{\sigma}$ & 50.0\\
        $\phi_i$ & $188.014$\\
        $\chi_i$ & $93.000$\\
            $\sigma_i$ & $40.326$\\
    \end{tabular}
     \caption{Three-field noninteracting case I: parameters of the potential and initial conditions. $M_p$ is set to 1.}
             \label{Paras-2}
  \end{center}
\end{table}
\\
The results for this case are shown in Figs. \ref{FieldsVSNe-3RTST}-\ref{FieldTraj3d-3RTST}. Figure \ref{FieldsVSNe-3RTST} shows the evolution of fields during inflation. Figure \ref{HepsVSNe-3RTS} shows the evolution of the Hubble and slow-roll parameters. The trajectory in the field space is also shown in the 2D point of view in Fig. \ref{FieldTraj2d-3RTST}, and in the 3D point of view in Fig. \ref{FieldTraj3d-3RTST}. As is obvious, there are two turns in the trajectory.

\subsubsection{Perturbative level}
\begin{figure}
     \centering     \includegraphics[width=\textwidth]{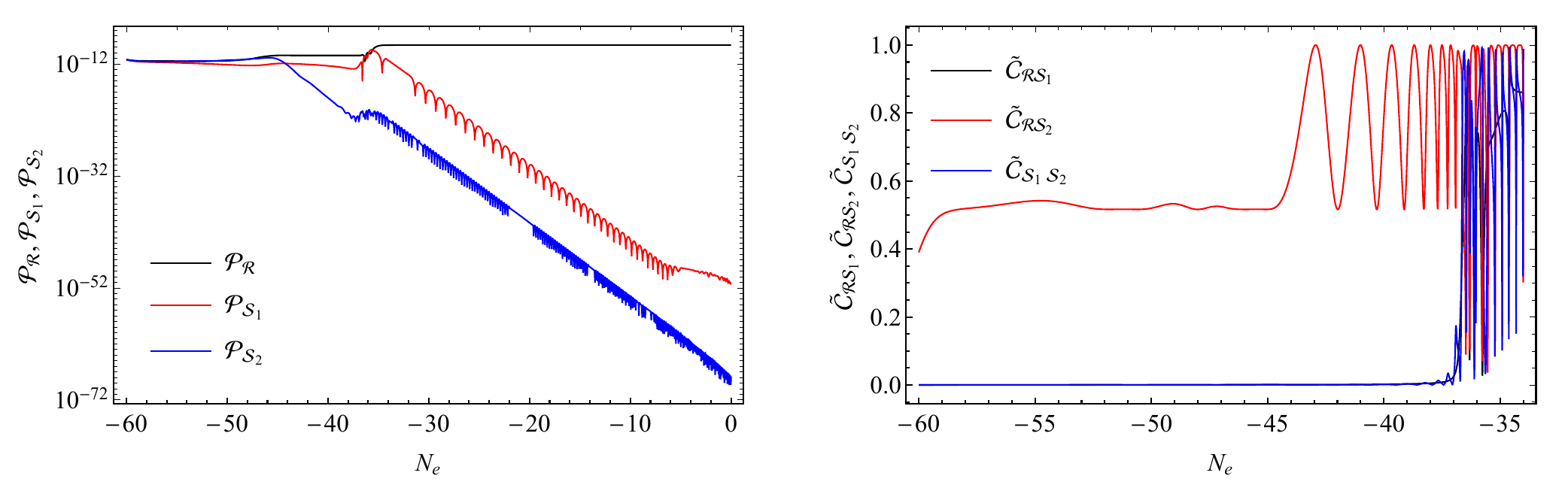}
     \caption{Three-field noninteracting case I: the evolution of the power spectra (left plot) of the curvature and isocurvature modes, and their correlations (right plot), defined in \eqref{Cij}, for the specific mode that exits the horizon 60 e-folds before the end of inflation, $ k= 0.002  \hspace{0.1cm} {\rm Mpc}^{-1}$, with respect to $N_e$. As it can be seen from the field space trajectory, shown in Fig. \ref{FieldTraj3d-3RTST}, there are two turns in the trajectory. The rates of these turns with respect to the Hubble rate are shown in the left plot of Fig. \ref{PS-TR-ns-3RTST}. The power spectrum of the curvature mode experiences a mild increase at the turns. The power spectra of isocurvature modes also, respectively, experience an increase at the two turns initially and then fall off rapidly along with oscillations. $\tilde{\mathcal{C}}_{\mathcal{R} \mathcal{S}_1}$ and  $\tilde{\mathcal{C}}_{\mathcal{S}_1 \mathcal{S}_2}$ have a similar behavior. Both correlations shift from zero to one at about the e-fold in which the second turn happens. $\tilde{\mathcal{C}}_{\mathcal{R} \mathcal{S}_2}$ shifts from zero to one at about the time where the first turn happens. After about 35 e-folds before the end of inflation, there is no special feature in the correlations, and there is just an oscillatory behavior due to the very small value and the oscillatory behavior of the power spectra of the isocurvature modes, which show themselves in the denominator of the correlation definition. See Eq. \eqref{Cij}}.
     \label{PS-Corr-Ne-3RTST}
 \end{figure}

\begin{figure}
     \centering     \includegraphics[width=\textwidth]{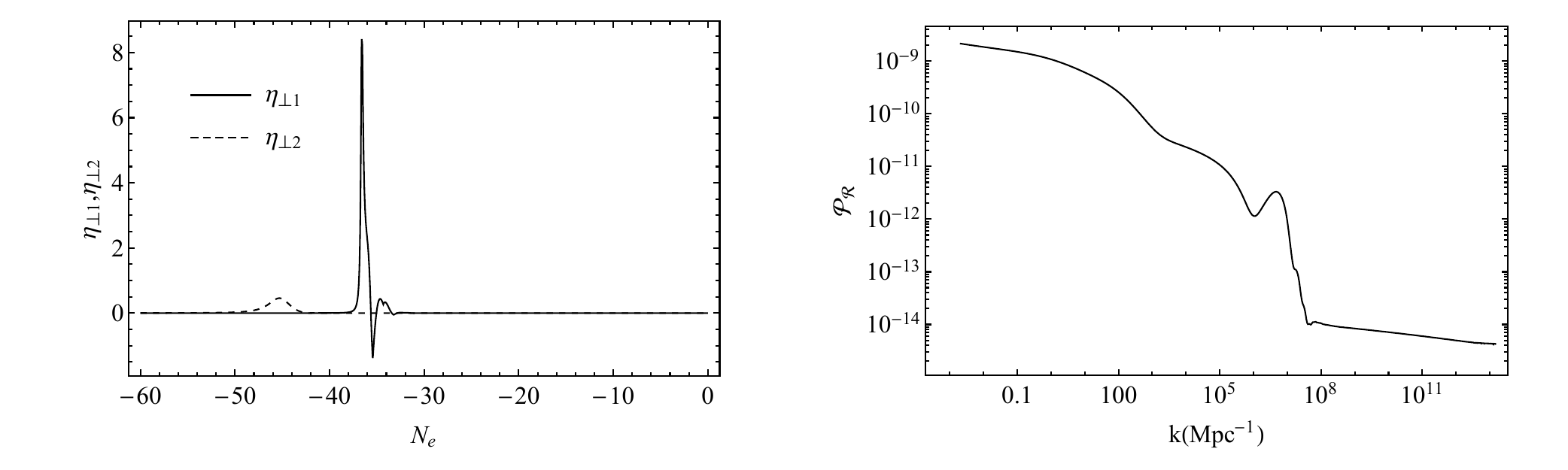}
     \caption{\textbf{Three-field noninteracting case I: turn rates of the field trajectory, defined in Eqs. \eqref{TurnRate1} and \eqref{TurnRate2}, are shown in the left plot. The first turn is a slow turn, which means that the turn rate is lower than one, and the second turn is a sharp turn. The power spectrum of the curvature mode for different values of $k$ evaluated at the end of inflation is shown in the right plot. Power spectra of the isocurvature modes could not be obtained numerically without noise, since their values are very small (respectively, of order $10^{-52}$ and $10^{-68}$). However, their general shape is similar to the power spectrum of the curvature mode.}}
     \label{PS-TR-ns-3RTST}
 \end{figure}

\begin{figure}
     \centering     \includegraphics[width=\textwidth]{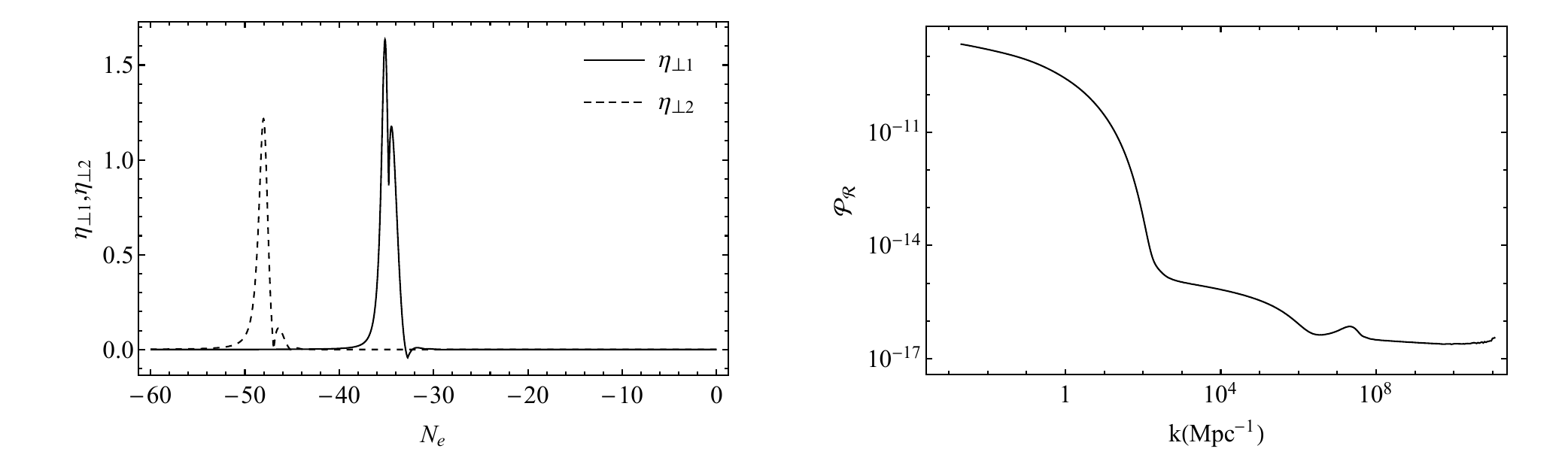}
     \caption{Three-field noninteracting case II: turning rates of the field trajectory, defined in Eqs. \eqref{TurnRate1} and \eqref{TurnRate2}, are shown in the left plot. As is obvious, one of the turning rates is about unity, and the other one is larger than one. The power spectra of the curvature mode for different values of $k$ evaluated at the end of inflation are shown in the right plot.}
     \label{PS-TR-ns-32RT}
 \end{figure}

\begin{figure}
     \centering     \includegraphics[width=\textwidth]{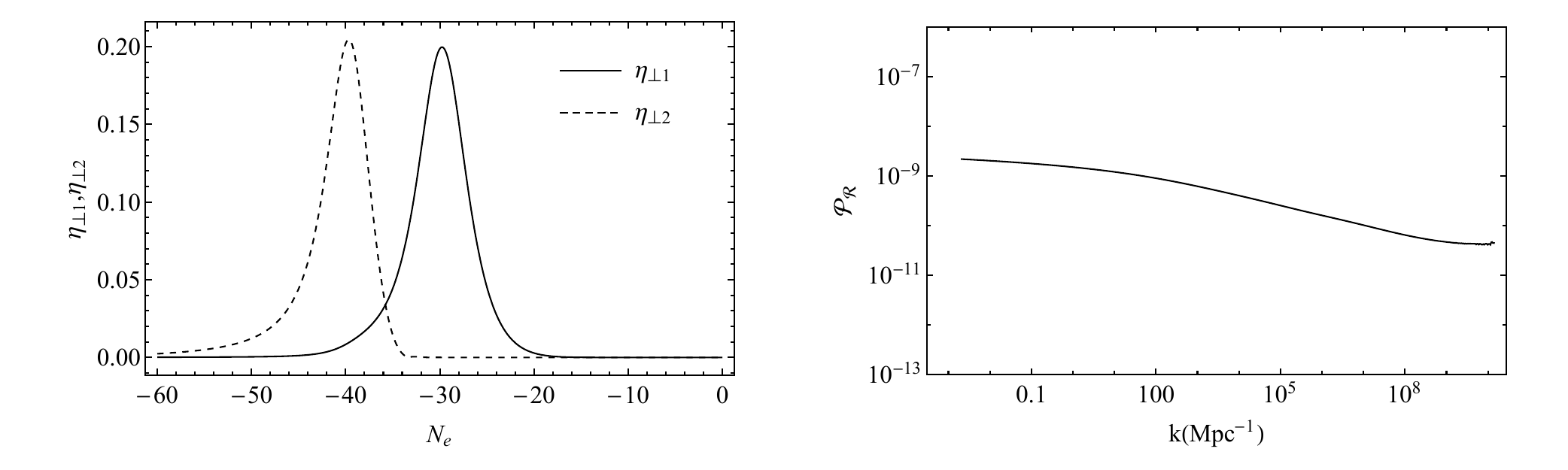}
     \caption{Three-field noninteracting case III: turning rates of the field trajectory, defined in Eqs. \eqref{TurnRate1} and \eqref{TurnRate2}, are shown in the left plot. As is obvious, both are lower than unity, therefore, we have two slow turns. The power spectrum of the curvature mode for different values of $k$ evaluated at the end of inflation is shown in the right plot.}
     \label{PS-TR-ns-32ST}
 \end{figure}

In this case we should solve the full Eqs. \eqref{diffequQl}—\eqref{diffequS2}, since none of the coefficients vanish. In Fig. \ref{PS-Corr-Ne-3RTST}, the evolution of the power spectra of the curvature and isocurvature modes (left plot), and the correlations (right plot), defined in eq. \eqref{Cij}, for a specific mode that exits the horizon 60 e-folds before the end of inflation, i.e., $ k= 0.002  \hspace{0.1cm}{\rm Mpc}^{-1}$, with respect to $N_e$ are shown. The field trajectory in this model has two turns at about 45  and 35 e-folds before the end of inflation. As can be seen in the left plot of Fig. \ref{PS-TR-ns-3RTST}, the first turn is a slow turn in the sense that $\eta_{\perp 1} < 1$, and the second turn is a rapid turn which means that $ \eta_{\perp 2} > 1$.  The power spectrum of the curvature mode experiences mild stepwise increases at the turns, and the power spectrums of the first and the second isocurvature modes first experience an increase, respectively, at the second and the first turns, and then begin a sharp decrease along with oscillations. The correlations between the first isocurvature mode and the curvature mode, $\tilde{\mathcal{C}}_{\mathcal{R} \mathcal{S}_1}$ (black line), and  between the two isocurvature modes, $\tilde{\mathcal{C}}_{\mathcal{S}_1 \mathcal{S}_2}$ (blue line), have almost a similar behavior. Both correlations shift from zero to one at about the e-fold in which the second turn happens. The correlations between the second isocurvature mode and the curvature mode, $\tilde{\mathcal{C}}_{\mathcal{R} \mathcal{S}_2}$ (red line), shift from zero to one at about the time where the first turn happens. After about 35 e-folds before the end of inflation, there is no special feature in the correlations, and there is just an oscillatory behavior due to the very small and oscillatory behavior of the power spectra of the isocurvature modes, which imprint themselves in the denominator of the correlation definition [see Eq. \eqref{Cij}].
Dependence of the power spectrum of the curvature mode at the end of inflation on \textit{k}, and the turning rates $\eta_{\perp 1}$ and $\eta_{\perp 2}$, are also shown in Fig. \ref{PS-TR-ns-3RTST}. The amplitudes of the power spectra of the isocurvature modes at the end of inflation are too small (respectively, of order $10^{-52}$ and $10^{-68}$). Therefore, the numerical results were dominated with noise. However, their general shapes, as much as we manage to investigate them, are similar to that of the curvature mode. The effect of the turns in the trajectory can be seen for the modes tht exit the horizon around the time the turns occur. The difference between the effect of the slow turn, happening at modes around $k = 10^3 \rm Mpc^{-1}$, and the rapid turn, happening at modes around $k = 10^6 \rm Mpc^{-1}$, is obvious in the right plot of Fig. \ref{PS-TR-ns-3RTST}. \\
By changing the potential parameters, different behaviors of the turning regimes in the field trajectory can be obtained. For instance we have obtained two other cases. The first case has two turns, one of which has a turning rate almost equal to one and the other one with a turn larger than unity. The parameters of this case are shown in Table \ref{Paras-32RT}, and the turning rates and power spectrum of the curvature mode are shown in Fig. \ref{PS-TR-ns-32RT}. The other case is a case with two slow turns. The parameters of this case are shown in Table \ref{Paras-32ST}, and the turning rates and power spectrum of the curvature mode are shown in Fig. \ref{PS-TR-ns-32ST}.

 \begin{table}[h!]
  \begin{center}
    \begin{tabular}{l|c|r} 
      \textbf{Parameters} & \textbf{Three-Field Noninteracting Case II}\\
      \hline
      &\\
      $\lambda_{\phi}$ & $ 0.950 \times 10^{-22}$\\
      $\lambda_{\chi}$ & $90  \lambda_{\phi} \left(\frac{\mu_{\phi}}{\mu_{\chi}}\right)^2$  \\
       $\lambda_{\sigma}$ & $2400 \lambda_{\phi} \left(\frac{\mu_{\phi}}{\mu_{\sigma}}\right)^2$  \\
      $\mu_{\phi}$ & 200.0\\
        $\mu_{\chi}$ & 100.0\\
         $\mu_{\sigma}$ & 50.0\\
        $\phi_i$ & $188.014$\\
        $\chi_i$ & $93.000$\\
            $\sigma_i$ & $40.326$\\
    \end{tabular}
    \caption{Three-field noninteracting case II: parameters of the potential and initial conditions of the case that have a turn with a turning rate almost equal to one and a rapid turn. $M_p$ is set to 1.}
             \label{Paras-32RT}
  \end{center}
\end{table}

\begin{table}[h!]
  \begin{center}
    \begin{tabular}{l|c|r} 
      \textbf{Parameters} & \textbf{Three-Field Noninteracting Case III}\\
      \hline
        & \\
      $\lambda_{\phi}$ & $ 0.140 \times 10^{-15}$\\
      $\lambda_{\chi}$ & $20  \lambda_{\phi} \left(\frac{\mu_{\phi}}{\mu_{\chi}}\right)^2$  \\
       $\lambda_{\sigma}$ & $100 \lambda_{\phi} \left(\frac{\mu_{\phi}}{\mu_{\sigma}}\right)^2$  \\
      $\mu_{\phi}$ & 200.0\\
        $\mu_{\chi}$ & 100.0\\
         $\mu_{\sigma}$ & 50.0\\
        $\phi_i$ & $187.920$\\
        $\chi_i$ & $93.000$\\
            $\sigma_i$ & $40.326$\\
    \end{tabular}
     \caption{Three-field noninteracting case III: parameters of the potential and initial conditions of a case having two slow turns. $M_p$ is set to 1.}
             \label{Paras-32ST}
  \end{center}
\end{table}

\subsection{Three-Field Interacting Case}
After investigating the spectrum of a three-field inflationary model without any direct interaction between different fields, now we are turning to the three-field case involving direct interactions between different fields. Therefore, the potential of this case has an interacting term, $V_{\phi\chi\sigma}$, besides the previous terms, Eqs. \eqref{potentialphic2}—\eqref{potentialsigc2},
\begin{align}
V_{\phi\chi\sigma} =  g(\phi^2 \chi^2 + \phi^2 \sigma^2 + \chi^2 \sigma^2).
    \label{interactingpot}
\end{align}

\subsubsection{Background solution}

\begin{figure}
     \centering
     \includegraphics[width=\textwidth]{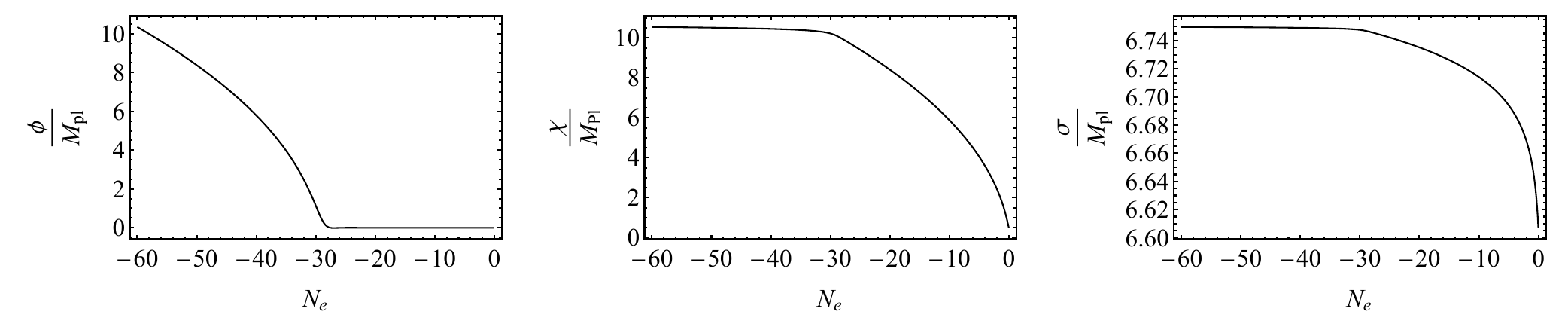}
\caption{Three-field interacting case: the evolution of the three scalar fields, $\phi$, $\chi$, and $\sigma$, is shown with respect to the number of e-folds. $N_e=0$ is the end of inflation}
     \label{FieldsVSNe-3Int}
 \end{figure}

\begin{figure}
     \centering
     \includegraphics[width=\textwidth]{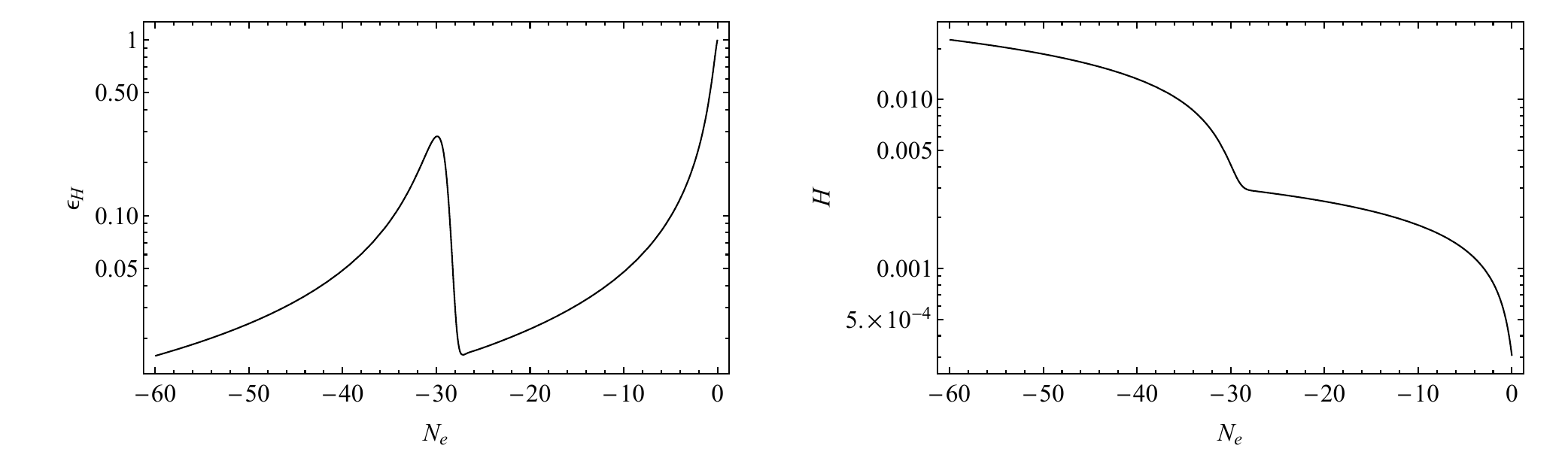}
\caption{\textbf{Three-field interacting case: the evolution of the Hubble parameter, \textit{H}, and the slow-roll parameter, $\epsilon_H$, is shown.}}
\label{HepsNe-3Int}
 \end{figure}

\begin{figure}
     \centering
     \includegraphics[width=\textwidth]{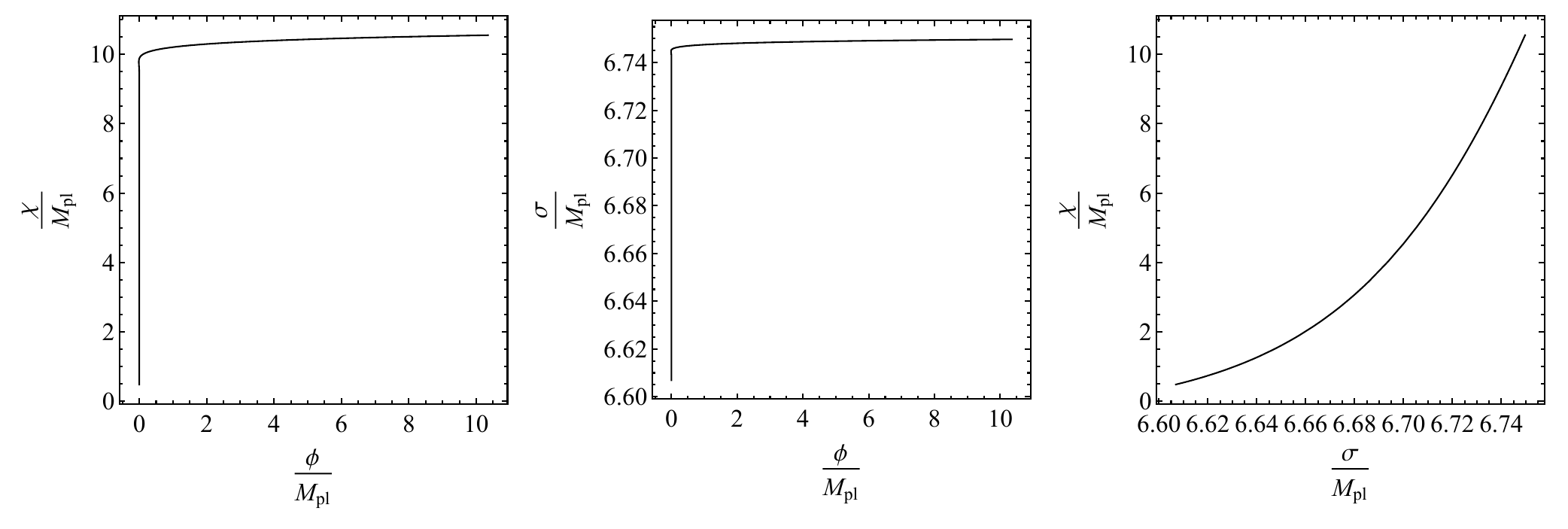}
\caption{Three-field interacting case: the 2D projections of the background trajectory in the field space.}
     \label{FieldTraj2d-3Int}
 \end{figure}

\begin{figure}
     \centering
     \includegraphics[width=0.5\textwidth]{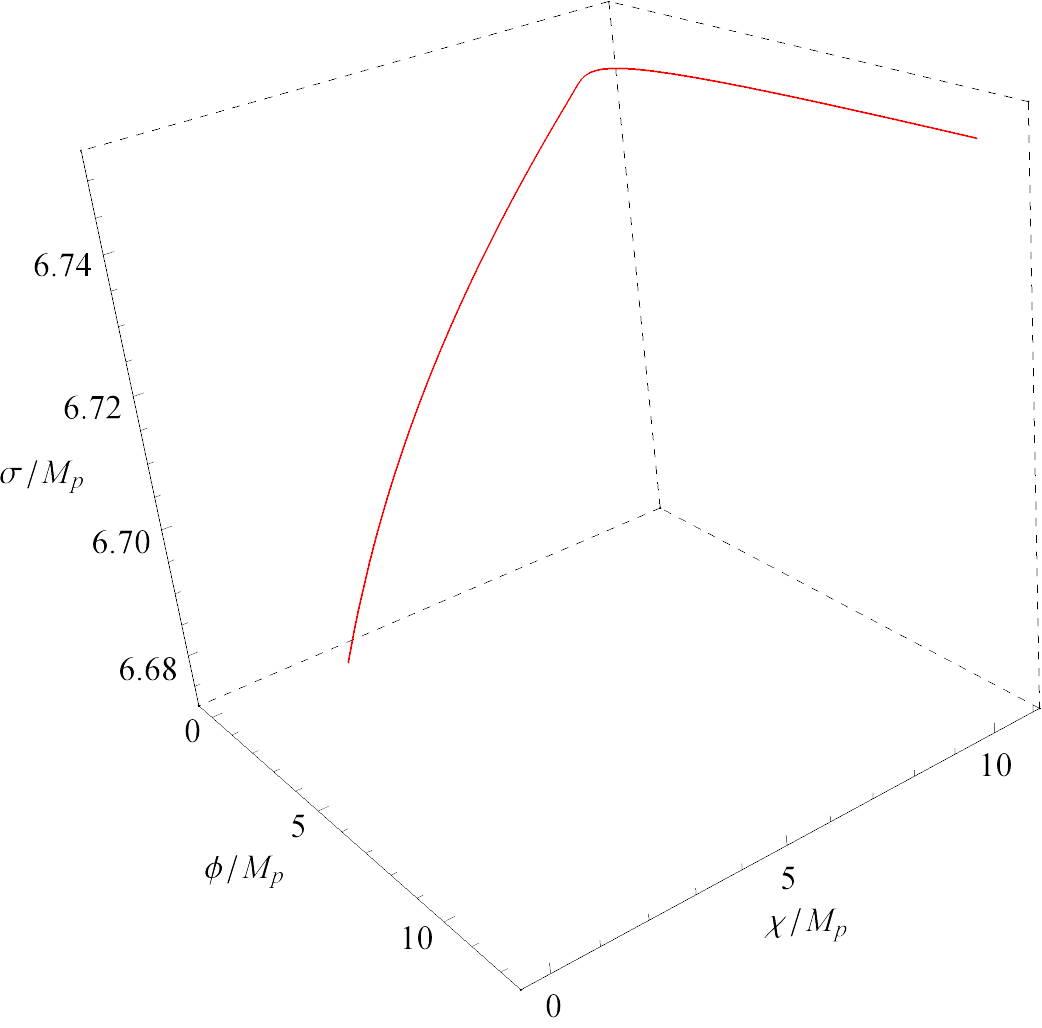}
\caption{\textbf{Three-field interacting case: the 3D background trajectory in the field space.}}
     \label{FieldTraj3d-3Int}
 \end{figure}

Parameters and initial conditions of this case are shown in Table \ref{Paras-3Int}.
The results for this case are shown in Figs. \ref{FieldsVSNe-3Int}—\ref{FieldTraj3d-3Int}. Figure \ref{FieldsVSNe-3Int} shows the evolution of fields during inflation. Figure \ref{HepsNe-3Int} shows the evolution of the Hubble and slow-roll parameters. The trajectory in the field space in the 2D point of view is shown in Fig. \ref{FieldTraj2d-3Int}, and from a 3D point of view in Fig. \ref{FieldTraj3d-3Int}.
\begin{table}[h!]
  \begin{center}
    \begin{tabular}{l|c|r} 
      \textbf{Parameters} & \textbf{Three-Field Interacting Case}\\
      \hline
      &\\
      $\lambda_{\phi}$ & $4500  \lambda_{\sigma} \left(\frac{\mu_{\sigma}}{\mu_{\phi}}\right)^2$ \\
      $\lambda_{\chi}$ & $90  \lambda_{\sigma} \left(\frac{\mu_{\sigma}}{\mu_{\chi}}\right)^2$  \\
      $\lambda_{\sigma}$ & $ 5.520 \times 10^{-12}$\\
       $g$ & $5.000 \times 10^{-22}$\\
      $\mu_{\phi}$ & 200.00\\
        $\mu_{\chi}$ & 100.00\\
         $\mu_{\sigma}$ & 50.00\\
        $\phi_i$ & $13.441$\\
        $\chi_i$ & $10.600$\\
            $\sigma_i$ & $6.750$\\
    \end{tabular}
     \caption{Three-field interacting case: parameters of the potential and initial conditions for the interacting case. $M_p$ is set to 1.}
\label{Paras-3Int}
\end{center}
\end{table}

\subsubsection{Perturbative level}
\begin{figure}
     \centering
     \includegraphics[width=\textwidth]{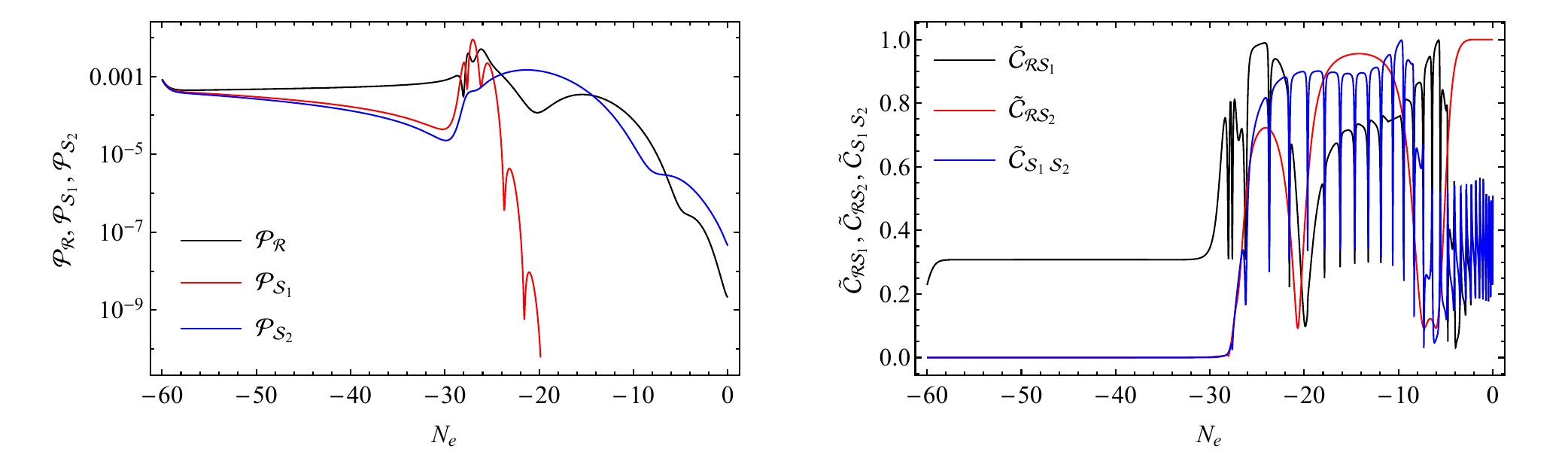}
     \caption{Three-field interacting case: the evolution of the power spectra of the curvature and isocurvature modes for a specific Fourier mode, which exits the horizon 60 e-folds before the end of inflation,  $k= 0.002  \hspace{0.1cm} {\rm Mpc}^{-1}$, with respect to $N_e$. At about 30 e-folds before the end of inflation, the power spectra of curvature and isocurvature modes experience a jump and then fall off. The power spectrum of the second isocurvature mode falls off more slowly, therefore, its amplitude at the end of inflation is larger than the power spectrum of the curvature mode. $\tilde{\mathcal{C}}_{\mathcal{R} \mathcal{S}_1}$ is almost constant until when the turn happens. The second isocurvature mode just becomes correlated with the curvature and the first isocurvature modes when the turn happens, see $\tilde{\mathcal{C}}_{\mathcal{S}_1 \mathcal{S}_2}$ and $\tilde{\mathcal{C}}_{\mathcal{R} \mathcal{S}_2}$.}
     \label{PS-Corr-Ne-3Int}
 \end{figure}

Solving Eqs. \eqref{diffequQl}—\eqref{diffequS2} numerically, we obtained the desired quantities. The evolution of the power spectra of the curvature and isocurvature modes (left plot), and their correlations (right plot), for a specific mode, which exits the horizon 60 e-folds before the end of inflation, $ k= 0.002  \hspace{0.1cm} {\rm Mpc}^{-1}$, with respect to $N_e$ are shown in Fig. \ref{PS-Corr-Ne-3Int}. The field trajectory in this case has a sharp turn at about 30 e-folds before the end of inflation. The power spectra of the curvature and isocurvature experience jumps at the turn. After that the power spectra of the curvature and first isocurvature mode start to fall off until the end of inflation to, respectively, reach values of order $ \mathcal{P}_{\mathcal{R}} \simeq 10^{-9}$, and $\mathcal{P}_{s_1} \simeq 10^{-39}$. The power spectrum of the second isocurvature mode falls off more slowly, and its amplitude at the end of inflation is larger than the curvature mode ($ \mathcal{P}_{s_2} \simeq 10^{-8}$). The correlations between the first isocurvature mode and the curvature mode, $\tilde{\mathcal{C}}_{\mathcal{R} \mathcal{S}_1}$ (black line), have an almost constant value of 0.3 until the turn happens at about 30 e-folds before the end of inflation, and begins to oscillate between 0.3 and 1. The second isocurvature mode is uncorrelated to the curvature and the first isocurvature mode ($\tilde{\mathcal{C}}_{\mathcal{S}_1 \mathcal{S}_2}$, $\tilde{\mathcal{C}}_{\mathcal{R} \mathcal{S}_2}$) until when the turn happens.
 
\begin{figure}
     \centering     
     \includegraphics[width=\textwidth]{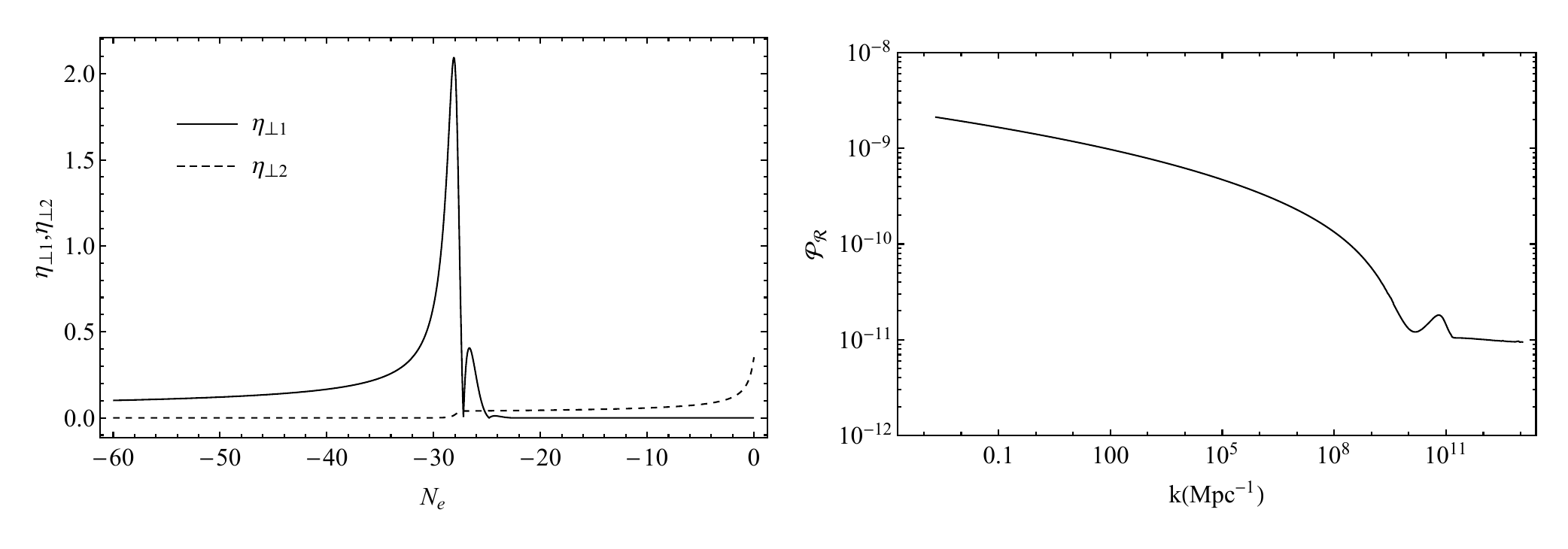}
     \caption{Three-field interacting case: in the left plot, the turning rates of the trajectory are shown. $\eta_{\perp 2}$ is negligible during the inflation, and there is only a single rapid turn in this case.
    The power spectrum of the curvature mode at the end of inflation for different momentum modes is shown in the right plot. The value of the power spectrum for the mode that exits the horizon at the CMB scale is set to the known $2.1 \times 10^{-9}$. The power spectra of the isocurvature modes have similar shapes with different values, $ \mathcal{P}_{s_1} \sim 10^{-39}$ and $ \mathcal{P}_{s_2} \sim 10^{-7}$.}
     \label{PS-TR-ns-3Int}
 \end{figure}

In Fig. \ref{PS-TR-ns-3Int}, the power spectrum of the curvature mode is shown with respect to $k$ (right plot).  The value of the power spectrum for the mode that exits the horizon at the CMB scale is set to the known value, $2.1 \times 10^{-9}$. The power spectra of the isocurvature modes have similar shapes with amplitudes $ \mathcal{P}_{s_1} \sim 10^{-39}$ and $ \mathcal{P}_{s_2} \sim 10^{-7}$ for $k=0.002 \hspace{0.1cm} \rm Mpc^{-1}$. The turn rates $\eta_{\perp 1}$ and $\eta_{\perp 2}$ are also shown in the same figure (left plot).

\section{Relation to Isocurvature Perturbations During  Radiation-Dominated Era}\label{RTO}

The large-scale structures we observe today originated from primordial cosmological perturbations during the radiation-dominated era, around the epoch of primordial nucleosynthesis ($T \sim 10^{10}$ K). These primordial perturbations are typically assumed to be purely adiabatic. While this assumption is consistent with key observations such as the CMB, it remains valid to consider initial conditions that include a subdominant isocurvature component. In general, one can consider four distinct types of isocurvature perturbations: the baryon isocurvature mode, the cold dark matter isocurvature mode, the neutrino isocurvature mode, and the neutrino velocity isocurvature mode\cite{Bucher:1999re}.

The composition of the initial perturbations is determined by the evolution of the fields during inflation and the subsequent reheating process. For isocurvature perturbations to survive, they must remain decoupled from the thermal bath and avoid thermalization throughout the postinflationary history of the Universe \cite{Polarski:1994rz}. In general, if the fields and their decay products thermalize completely after inflation, no entropy perturbations can persist in a thermal equilibrium state characterized by a single temperature. This implies that a neutrino isocurvature perturbation is unlikely to be produced by inflation unless the reheating temperature is comparable to the neutrino decoupling temperature shortly before primordial nucleosynthesis \cite{Lyth:1998xn,Weinberg:2008zzc}. In contrast, a cold dark matter species can easily remain decoupled at such temperatures. Consequently, the simplest and most plausible scenario for the presence of isocurvature modes in the early radiation-dominated era is the cold dark matter isocurvature mode \cite{Amendola:2001ni}.

In addition to the four types of isocurvature perturbations mentioned before, the existence of new light relics, or dark radiation, is a well-motivated extension of the standard model, with implications for both fundamental particle physics and cosmological observables. Such particles, if thermally coupled and subsequently decoupled in the early Universe, contribute to the total relativistic energy density. This is conventionally parametrized by the effective number of neutrino species, $N_{\rm eff}$. While the standard model provides a precise prediction of $N_{\rm eff}\approx 3.044$ \cite{deSalas:2016ztq,Mangano:2005cc, Bennett:2020zkv}, current constraints from the CMB and big bang nucleosynthesis allow for a small excess, $\Delta N_{\rm eff} \gtrsim 0.2-0.3$ \cite{Aghanim:2018eyx}, which could be explained by such new physics. This framework encompasses a broad spectrum of candidates, including QCD axions \cite{Wantz:2010it}, sterile neutrinos \cite{Dodelson:1993je, Hamann:2011ge, Mirizzi:2012we, Aghanim:2018eyx}, and light particles from a thermally decoupled dark sector \cite{Cyburt:2015mya}. Since these components decouple at different times or through different interactions than standard model neutrinos, they can maintain independent perturbation histories. Consequently, dark radiation provides a compelling and well-motivated origin for (nonadiabatic) isocurvature initial conditions, offering a potential link between particle physics beyond the standard model and the primordial perturbation spectrum (see, e.g., \cite{Kawasaki:2011rc, Kawakami:2012ke,Buckley:2024nen}).

The presence of three fields during inflation leads to a richer, though more complex, phenomenology of isocurvature modes. To facilitate a phenomenological comparison between two-field and three-field models, we examine a universe with two initial isocurvature perturbations, $\mathcal{S}_C$ and $\mathcal{S}_{DR}$, in the early radiation-dominated era. We analyze this scenario first for an $N$-field inflationary model and then under the assumption of a two-field, and subsequently a three-field inflationary model. The following analysis is based on a scenario in which inflationary dynamics generate isocurvature modes for both dark matter and dark radiation, ~\footnote{One could alternatively consider dominant and subdominant dark matter components with distinct properties. In this case, in the following argument with minor modifications, ``$C$'' would refer to the dominant dark matter component while ``$DR$'' would refer to the subdominant dark matter component.}
\begin{align}
    \mathcal{S}_{C} &= \frac{\delta \rho_c}{\rho_c} - \frac{3}{4}\frac{\delta \rho_{\gamma}}{\rho_{\gamma}}, \\
    \mathcal{S}_{DR} &= \frac{3}{4}\Big(\frac{\delta \rho_{DR}}{\rho_{DR}} - \frac{\delta \rho_{\gamma}}{\rho_{\gamma}}\Big).
\end{align}
We show the curvature and isocurvature modes at the early domination era by a $3 \times 1$ matrix $ R_r $ as
\begin{align}
    R_r = \begin{pmatrix}
        \mathcal{R}_R \\ \mathcal{S}_C \\ \mathcal{S}_{DR}
    \end{pmatrix}.
\end{align}
The same matrix for the modes in the inflationary era, and at the end of inflation, is showed by $R_I$. This matrix in genral is a $N \times 1$ matrix for an $N$-field inflationary model,
\begin{align}
    R_I = \begin{pmatrix}
        \mathcal{R}_I \\ \mathcal{S}_1 \\
        \vdots\\
        \mathcal{S}_{N-1}
    \end{pmatrix}.
\end{align}
The observable quantities are in fact the correlation functions of the modes in $R_r$, which we can put in the correlation matrix,
\begin{align}
    \mathcal{O}^R =  \begin{pmatrix}
         \mathcal{P}_{\mathcal{R}}& \mathcal{C}_{\mathcal{R}\mathcal{S}_C}& \mathcal{C}_{\mathcal{R}\mathcal{S}_{DR}} \\
       \mathcal{C}_{\mathcal{R}\mathcal{S}_C}& \mathcal{P}_{\mathcal{S}_C}& \mathcal{C}_{\mathcal{S_C} \mathcal{S}_{DR}}\\
\mathcal{C}_{\mathcal{R}\mathcal{S}_{DR}}& \mathcal{C}_{\mathcal{S_C} \mathcal{S}_{DR}}& \mathcal{P}_{\mathcal{S}_{DR}}
     \end{pmatrix}.
\end{align}
Note that this matrix is symmetric and has in general, six independent correlators. $R_r$ is in general related to $R_I$ by the transfer matrix, $T$, which is a $ 3 \times N $ matrix,
\begin{align}
    R_r = T R_I\,.
\end{align}
Physical considerations impose constraints on the general form of the transfer matrix. First, if no primordial isocurvature perturbations exist at the end of inflation, none can be produced subsequently, and the adiabatic mode must remain constant. This implies the conditions $T_{\mathcal{R}\mathcal{R}} = 1$, and $T_{\mathcal{S}_C\mathcal{R}} = T_{\mathcal{S}_{DR}\mathcal{R}} = 0$. The dynamics of the perturbations after horizon exit are thus determined by the remaining elements of the transfer matrix, whose specific forms depend on the microphysical model,
\begin{align}\label{TranFunc}
   T = \begin{pmatrix}
         1& T_{\mathcal{R}\mathcal{S}_1}& \dots &  T_{\mathcal{R}\mathcal{S}_{N-1}}\\
         0& T_{\mathcal{S}_C\mathcal{S}_1}& \dots &T_{\mathcal{S}_C\mathcal{S}_{N-1}} \\
         0& T_{\mathcal{S}_{DR}\mathcal{S}_1}& \dots &T_{\mathcal{S}_{DR}\mathcal{S}_{N-1}}
     \end{pmatrix}_{3 \times N}
\end{align}
The two-point correlation matrix in the radiation-dominated phase, $\mathcal{O}^R$, can be obtained from the inflationary two-point correlation function and the transfer matrix,
\begin{align}\label{OR-OI-T}
      \mathcal{O}^R =  T\mathcal{O}^I T^{\intercal}\,,
\end{align}
where
\begin{align}
      \mathcal{O}^I = \langle R_I R_I^{\intercal} \rangle\,,
\end{align}
In general, for an $N$-field model of inflation, we have $\frac{N(N+1)}{2}$ independent correlation functions in $\mathcal{O}^I$.  The six independent correlation functions of $\mathcal{O}^R$ are related linearly to these inflationary two-point correlation functions in $\mathcal{O}^I$. We can extract the following conclusions \cite{Horn:2012}:
\begin{itemize}
\item In a single-field model, the rank of the correlation matrix, $\mathcal{O }^R$, can be at most 1. Therefore, there are consistency relations for single-field models stating that the
determinant of $\mathcal{O}^R$ and the determinants of $2\times2$ submatrices of $\mathcal{O}^R$ should vanish.
    \item In a two-field model, the rank of the correlation matrix, $\mathcal{O}^R$, can be at most 2. Therefore, there is a consistency relation for two-field models stating that the determinant of $\mathcal{O}^R$ should vanish.
    \item In a three-field model, the rank of the correlation matrix, $\mathcal{O}^R$, can be at most 3.
    \item In an $N$-field model, with $N \geq 4$, the rank of  the correlation matrix, $\mathcal{O}^R$, can be again at most 3.
    \end{itemize}
The preceding arguments demonstrate that three-field inflationary models yield a richer phenomenology for generating two isocurvature modes in the early radiation-dominated era than their two-field counterparts. This framework provides a method for establishing a lower bound on the number of degrees of freedom active during inflation. Specifically, if observations indicate that the correlation matrix $\mathcal{O}^R$ from the early radiation era has a rank of 3, it would imply the involvement of at least three independent degrees of freedom during inflation.

The enhanced phenomenological freedom of three-field inflation, compared to the two-field case, when two initial isocurvature modes are required, can be understood from a parameter-counting perspective too. The components of the correlation matrix can be written explicitly in terms of the two-point correlation functions of a general $N$-field case as
\begin{align}
\label{PRr}
    \mathcal{P}_{\mathcal{R}} &= \mathcal{P}_{\mathcal{R}}^* + \sum_{J=1}^{N-1} \mathcal{P}_{\mathcal{S}_J}T_{\mathcal{R}\mathcal{S}_J}^2 +  2\sum_{J=1}^{N-1} \mathcal{C}^*_{\mathcal{R}\mathcal{S}_J}T_{\mathcal{R}\mathcal{S}_J} + \sum_{J=1}^{N-1}\sum_{K=1}^{N-1} \delta_{JK}\mathcal{C}^*_{\mathcal{S}_J\mathcal{S}_K}T_{\mathcal{R}\mathcal{S}_J}T_{\mathcal{R}\mathcal{S}_K} \\
    \label{PSCr}
     \mathcal{P}_{\mathcal{S}_C} &=  \sum_{J=1}^{N-1} \mathcal{P}^*_{\mathcal{S}_J}T_{\mathcal{S}_C\mathcal{S}_J}^2  + \sum_{J=1}^{N-1}\sum_{K=1}^{N-1} \delta_{JK}\mathcal{C}^*_{\mathcal{S}_J\mathcal{S}_K}T_{\mathcal{S}_C\mathcal{S}_J}T_{\mathcal{S}_C\mathcal{S}_K}  \\
     \label{PSDRr}
     \mathcal{P}_{\mathcal{S}_{DR}} &=  \sum_{J=1}^{N-1} \mathcal{P}^*_{\mathcal{S}_J}T_{\mathcal{S}_{DR}\mathcal{S}_J}^2  + \sum_{J=1}^{N-1}\sum_{K=1}^{N-1} \delta_{JK}\mathcal{C}^*_{\mathcal{S}_J\mathcal{S}_K}T_{\mathcal{S}_{DR}\mathcal{S}_J}T_{\mathcal{S}_{DR}\mathcal{S}_K} \\
     \label{CRSCr}
\mathcal{C}_{\mathcal{R}\mathcal{S}_C} &= \sum_{J=1}^{N-1} \mathcal{P}^*_{\mathcal{S}_J}T_{\mathcal{R}\mathcal{S}_J}T_{\mathcal{S}_C\mathcal{S}_J} + \sum_{J=1}^{N-1} \mathcal{C}^*_{\mathcal{R}\mathcal{S}_J} T_{\mathcal{S}_C\mathcal{S}_J} + \sum_{J=1}^{N-1}\sum_{K=1}^{N-1} \delta_{JK}\mathcal{C}^*_{\mathcal{S}_J\mathcal{S}_K}T_{\mathcal{R}\mathcal{S}_J}T_{\mathcal{S}_C\mathcal{S}_K}   \\
\label{CRSDRr}
     \mathcal{C}_{\mathcal{R}\mathcal{S}_{DR}} &= \sum_{J=1}^{N-1} \mathcal{P}^*_{\mathcal{S}_J}T_{\mathcal{R}\mathcal{S}_J}T_{\mathcal{S}_{DR}\mathcal{S}_J} + \sum_{J=1}^{N-1} \mathcal{C}^*_{\mathcal{R}\mathcal{S}_J} T_{\mathcal{S}_{DR}\mathcal{S}_J} + \sum_{J=1}^{N-1}\sum_{K=1}^{N-1} \delta_{JK}\mathcal{C}^*_{\mathcal{S}_J\mathcal{S}_K}T_{\mathcal{R}\mathcal{S}_J}T_{\mathcal{S}_{DR}\mathcal{S}_K}
     \\
     \label{CSCSDRr}
     \mathcal{C}_{\mathcal{S}_C\mathcal{S}_{DR}} &= \sum_{J=1}^{N-1} \mathcal{P}^*_{\mathcal{S}_J}T_{\mathcal{S}_C\mathcal{S}_J}T_{\mathcal{S}_{DR}\mathcal{S}_J} +\sum_{J=1}^{N-1}\sum_{K=1}^{N-1} \delta_{JK}\mathcal{C}^*_{\mathcal{S}_J\mathcal{S}_K}T_{\mathcal{S}_C\mathcal{S}_J}T_{\mathcal{S}_{DR}\mathcal{S}_K}
\end{align}
\\
 As previously established, the correlation matrix for perturbations in the radiation-dominated era, $\mathcal{O}^R$, possesses six independent elements. In a two-field inflationary model, there are six unknowns corresponding to three independent correlation functions at the end of inflation and three degrees of freedom in the transfer matrix [see Eq. \eqref{TranFunc}]. However, since the rank of $\mathcal{O}^R$ is two, its determinant is zero. This reduces the number of independent equations to at most five. Thus, the system is underconstrained, and we are still left with one undetermined degree of freedom. For the case where none of the free parameters in $T$ and $\mathcal{O}^R$ are equal to zero, one can show that a two-parameter family of solutions for $T$ and $\mathcal{O}^I$ exists. Besides the determinant constraint, one of the following consistency conditions should then hold for such a solution:
 \begin{align}
 &\mathcal{P}_{\mathcal{S}_C} \mathcal{P}_{\mathcal{S}_{DR}}=\mathcal{C}_{\mathcal{S_C} \mathcal{S}_{DR}}^2 \label{first-consistency}\,,\\
&\mathcal{C}_{\mathcal{R}\mathcal{S}_C} \mathcal{C}_{\mathcal{S_C} \mathcal{S}_{DR}}=\mathcal{C}_{\mathcal{R}\mathcal{S}_{DR}} \mathcal{P}_{\mathcal{S}_C} \label{second-consistency}\,.
 \end{align}
 Combining the above two equations, one can also verify that the determinant constraint is also satisfied. Therefore, instead of the determinant condition and one of the equations above, we proceed with the above two constraints.
 The first constraint, Eq. \eqref{first-consistency}, would mean that $\tilde{\mathcal{C}}^2_{\mathcal{S_C} \mathcal{S}_{DR}}=1$, which means that the dark matter and dark radiation entropy perturbations are either correlated or anticorrelated. Combining the second and first consistency relations, one obtains
 \begin{equation}\label{PDR-over-PSD}
    \frac{\mathcal{P}_{\mathcal{S}_{DR}}}{\mathcal{P}_{\mathcal{S}_C}}=\left(\frac{\mathcal{C}_{R \mathcal{S}_{DR}}}{\mathcal{C}_{\mathcal{R}\mathcal{S}_C}}\right)^2.
\end{equation}
One can also demonstrate that these two ratios are determined by the branching ratio of inflationary isocurvature perturbation, $\mathcal{S}_{1}$, to the dark radiation perturbation, over its branching ratio to the cold dark matter perturbation, $(T_{\mathcal{S}_{DR}\mathcal{S}_{1}}/T_{\mathcal{S}_C\mathcal{S}_{1}})^2$.

In a three-field model, the situation is similar. While the number of observables in $\mathcal{O}^R$ remains six, the number of theoretical parameters doubles: there are six independent correlation functions at the end of inflation and six independent components in the transfer matrix. This leaves six parameters underdetermined by the observations. Consequently, three-field models provide significantly greater freedom for modeling reheating and postreheating physics.

One can show that if the number of isocurvature modes during the radiation-dominated phase is $i$, and the rank of the corresponding two-point correlation matrix, $\mathcal{O}^R$, is $r \leq i+1$, then the number of independent components of $\mathcal{O}^R$ is given by \cite{Horn:2012}
\begin{equation}
r(i+1) - \frac{r(r-1)}{2}.
\end{equation}
This quantity must be smaller than the total number of degrees of freedom available in $\mathcal{O}^I$ for an $N$-field inflationary model, namely $\frac{N(N+1)}{2}$, plus the degrees of freedom of the transfer matrix, $(N-1)(i+1)$. This requirement sets a lower bound on the number of fields participating in inflation,
\begin{equation}\label{Nmin}
N \geq N_{\rm min} \equiv
\left\lceil
\frac{1}{2}
\left(
\sqrt{8 i r + 4 i (i+5) - 4 r^2 + 12 r + 17}
- 2i - 3
\right)
\right\rceil ,
\end{equation}
where $\lceil \cdot \rceil$ denotes the ceiling function.

For instance, when $i=4$ and $r=3$, one finds $N_{\rm min}=3$. In this case, the number of degrees of freedom of $\mathcal{O}^{R}$ is $10$, which is smaller than the combined total of independent components of $T$ and $\mathcal{O}^{I}$, namely, $14$. On the other hand, if $i=3$ and $r=2$, the argument of the ceiling function is an integer for $N=3$ and the equality holds in \eqref{Nmin}. This allows all independent components of $\mathcal{O}^I$ and $T$ to be determined uniquely in terms of the independent components of the correlation functions of the matrix $\mathcal{O}^R$. For larger values of $N$, the system remains underconstrained.

\section{Conclusion}\label{sec5}

In this work, we develop a three-field cosmological perturbation theory in flat field space. By introducing the so-called semikinematic  basis we derive explicit equations of motion for both the curvature perturbation and two distinct isocurvature modes, applicable to arbitrary potentials. We highlighted the characteristic property of more-than-two-field models, which is the freedom in choosing the directions of isocurvature perturbation. This framework facilitates efficient numerical computation of the mode evolution and their cross-correlations through the solutions of the resulting coupled linear differential equations. One should note that in comparison with the formalism developed in \cite{Pinol:2020kvw}, our formulation is computationally friendly and can be easily adjusted to tackle any potential.\\
As examples, three different scenarios have been investigated. First, we investigated a three-field model in which one of the fields remains at the bottom of the potential and which effectively reduces to a two-field model plus a linearly decoupled isocurvature mode. Our equations in the two-field limit are consistent with the two-field formulations in the literature. We later investigated two other scenarios of three-field inflation, where in the first case, different fields do not directly interact with each other, and in the second one, interactions between different fields are assumed. In the noninteracting case, we investigated three different subcases, which include different regimes of turns in the field space background trajectory. These subcases include slow turns, with turning rates lower than one, rapid turns, with turning rates larger than unity, and also marginal turns, with turning rates almost equal to one.

The effect of different turning regimes on the power spectra is shown. It is also observed that in all these cases, the amplitudes of the power spectra of isocurvature modes are quite subdominant. We then investigated a three-field model in which different fields directly interact with each other. In this case, we observed that the power spectrum of one of the isocurvature modes is about 3 orders of magnitude larger than the power spectrum of the curvature mode.

We also examined the observational implications of the three-field perturbation theory. A key finding is that the rank of the observable correlation matrix in the radiation-dominated phase sets a lower bound on the number of inflationary degrees of freedom in scenarios with two initial isocurvature modes. While in a two-field model at least one parameter (and at most two, if an additional consistency relation among the parameters of $\mathcal{O}^R$ holds) remains unconstrained, a three-field model introduces twelve unknown parameters—six from the transfer matrix and six from the correlation matrix. This indeterminacy provides at least six degrees of freedom for modeling reheating and the postinflationary universe, leading to a richer phenomenology.

In this paper, we worked in the flat field space. It would be interesting to explore generalizations of this work to models with a curved field space. Also, investigating the bispectrum when such sharp turns in the field space occur is another interesting avenue to pursue.



\acknowledgments

We would like to thank Abdulrahim Al Balushi for his collaboration in the initial stages of this project. We are also thankful to K. Turzynski for feedback on the manuscript. We are also grateful to M. M. Sheikh-Jabbari and M. H. Namjoo for useful discussions. This project is supported by INSF grant 4031449. This project has also received funding from the European Union's Horizon Europe research and innovation programme under the Marie Sk\L odowska-Curie Staff Exchange Grant Agreement No. 101086085 — ASYMMETRY. The work of S.M. was supported in part by the Japan Society for the Promotion of Science (JSPS) Grants-in-Aid for Scientific Research No.~24K07017 and the World Premier International Research Center Initiative (WPI), MEXT, Japan.

\section*{Data Availibility}
The data that support the findings of this article are openly available \cite{DataGit}

\appendix
\section{Transformation Matrices Derivation}
\label{Appendix-A}
In this appendix we show how the transformation matrices from the field space to the semikinematic and kinematic basis are obtained. We also show how we can transform between the semikinematic and kinematic basis.\\
As it is shown in Fig. \ref{traj-ins-basis} the adiabatic unit direction, and the field perturbation vector(FPV) are determined, respectively, by angles $\alpha$ and $\beta$,
\begin{align}
\label{sincosab}
\nonumber
    \sin \beta &= \frac{\sqrt{\dot{\phi}^2+\dot{\chi}^2}}{\dot{l}}, \hspace{2cm} \sin \alpha = \frac{\dot{\chi}}{\sqrt{\dot{\phi}^2+\dot{\chi}^2}},  \\
    \cos \beta &= \frac{\dot{\sigma}}{\dot{l}}, \hspace{3.4cm} \cos \alpha = \frac{\dot{\phi}}{\sqrt{\dot{\phi}^2+\dot{\chi}^2}},
\end{align}
and $\alpha'$ and $\beta'$,
\begin{align}
\nonumber
    \sin \beta' &= \frac{\sqrt{\delta\phi^2+\delta\chi^2}}{\sqrt{\delta\phi^2+\delta\chi^2 +\delta\sigma^2}}, \hspace{2cm} \sin \alpha' = \frac{\delta\chi}{\sqrt{\delta\phi^2+\delta\chi^2}},  \\
    \cos \beta' &= \frac{\delta\sigma}{\sqrt{\delta\phi^2+\delta\chi^2 +\delta\sigma^2}}, \hspace{2cm} \cos \alpha' = \frac{\delta\phi}{\sqrt{\delta\phi^2+\delta\chi^2}}.
\end{align}
Adiabatic unit vector and FPV can be written using the radial unit vector (just like $\hat{r}$ in spherical coordinates) as
\begin{align}
    &\hat{l} = \hat{\varphi}\sin \beta \cos \alpha + \hat{\chi}\sin \beta \sin \alpha +\hat{\sigma}\cos \beta, \\
    &\boldsymbol{\delta\Phi} = \sqrt{\delta\phi^2+\delta\chi^2 +\delta\sigma^2}(\hat{\varphi} \sin \beta' \cos \alpha' + \hat{\chi}\sin \beta' \sin \alpha' +\hat{\sigma}\cos \beta'),
\end{align}
therefore, the component of FPV along the adiabatic direction, $\delta l$, will be the inner product of $\hat{l}$, and $\boldsymbol{\delta\Phi}$
\begin{align}
\nonumber
   \delta l = \boldsymbol{\delta\Phi} \cdot \hat{l} =
   \begin{pmatrix}
       \delta \phi& \delta \chi& \delta \sigma
   \end{pmatrix}
   \begin{pmatrix}
       \sin \beta \cos \alpha \\
        \sin \beta \sin \alpha \\
        \cos \beta
   \end{pmatrix}
        = \delta\phi\sin \beta \cos \alpha + \delta\chi\sin \beta \sin \alpha +\cos \beta \delta \sigma.
\end{align}
 The two isocurvature directions we considered in the semikinematic basis are along the $\hat{\alpha_{l}}$ and $-\hat{\beta_{l}}$. In short, we have
\begin{align}
&\delta l=\boldsymbol{\delta\Phi} \cdot \hat{l} =\delta\phi\sin \beta \cos \alpha + \delta\chi\sin \beta \sin \alpha + \delta \sigma\cos \beta, \\
    &\delta s_1 =  \boldsymbol{\delta\Phi} \cdot \hat{\alpha}  = -\delta\phi \sin \alpha + \delta\chi \cos \alpha,  \\
     &\delta s_2 =  \boldsymbol{\delta\Phi} \cdot (-\hat{\beta})  = -\delta\phi\cos \beta \cos \alpha - \delta\chi\cos \beta \sin \alpha + \delta \sigma\sin \beta,
\end{align}
which can be written as \eqref{MSvstrans}—\eqref{Mdef}. Now, we want to obtain the transformation matrix from the field basis to the kinematic basis, $\mathcal{M}_k$. The unit adiabatic direction is defined as
\begin{align}
    \hat{l}^a = \frac{\dot{\Phi}^a}{\dot{l}}, \hspace{1cm} \dot{l} = \sqrt{\dot{\phi}^2 + \dot{\chi}^2 + \dot{\sigma}^2}.
\end{align}
The unit acceleration direction is defined as
\begin{align}
    \hat{n} = \frac{\dot{\hat{l}}}{|\dot{\hat{l}}|}
\end{align}
Using (\ref{BEql}), and \eqref{backequphi}—\eqref{backequsig},
we get
\begin{align}
    \dot{\hat{l}} &= \frac{\ddot{\Phi}^a}{\dot{l}} - \frac{\ddot{l}\dot{\Phi}^a}{\dot{l}^2} = \frac{1}{\dot{l}}(\hat{l}^aV_{,l}-\delta^{ab}V_{,b})
\end{align}
The amplitude of this vector is
\begin{align}
    |\dot{\hat{l}}|^2 &= \frac{1}{\dot{l}^2}\delta_{ab}(\hat{l}^a V_{,l}-\delta^{ac}V_{,c})(\hat{l}^b V_{,l}-\delta^{bd}V_{,d})
    = \frac{\delta^{cd}V_{,c}V_{,d}-V_{,l}^2}{\dot{l}^2}.
\end{align}
Therefore,
\begin{align}
    \nonumber
    \Rightarrow \hat{n}^a &= \frac{\hat{l}^aV_{,l}-\delta^{ab}V_{,b}}{(\delta^{cd}V_{,c}V_{,d}-V_{,l}^2)^{\frac{1}{2}}} \\
    &= \frac{1}{(\delta^{cd}V_{,c}V_{,d}-V_{,l}^2)^{\frac{1}{2}}}
    \begin{pmatrix}
      \frac{\dot{\phi}}{\dot{l}}V_{,l}-V_{,\phi}\\
       \frac{\dot{\chi}}{\dot{l}}V_{,l}-V_{,\chi}\\
     \frac{\dot{\sigma}}{\dot{l}}V_{,l}-V_{,\sigma}
\end{pmatrix} .
\end{align}
Having the above expression for the unit acceleration vector, we can calculate the isocurvature component along this direction at each point as
\begin{align}
\nonumber
    \delta s_n &= \boldsymbol{\delta\Phi} \cdot \hat{n} =  \frac{1}{(\delta^{cd}V_{,c}V_{,d}-V_{,l}^2)^{\frac{1}{2}}}
    \begin{pmatrix}
       \delta \phi& \delta \chi& \delta \sigma
    \end{pmatrix}
     \begin{pmatrix}
      \frac{\dot{\phi}}{\dot{l}}V_{,l}-V_{,\phi}\\
       \frac{\dot{\chi}}{\dot{l}}V_{,l}-V_{,\chi}\\
     \frac{\dot{\sigma}}{\dot{l}}V_{,l}-V_{,\sigma}
\end{pmatrix}\\
&=  \frac{1}{(\delta^{cd}V_{,c}V_{,d}-V_{,l}^2)^{\frac{1}{2}}}\Big[\Big(V_{,l}\sin \beta \cos \alpha-V_{,\phi}\Big)\delta\phi+\Big(V_{,l}\sin \beta \sin \alpha-V_{,\chi}\Big)\delta \chi+\Big(V_{,l}\cos \beta-V_{,\sigma}\Big)\delta \sigma\Big]
\end{align}
where we have used (\ref{sincosab}) in the second line. The third direction in the kinematic basis is naturally the cross product of the unit adiabatic and unit acceleration vectors
\begin{align}
  \hat{b} &= \hat{l} \times \hat{n} = \frac{1}{(\delta^{cd}V_{,c}V_{,d}-V_{,l}^2)^{\frac{1}{2}}}
    \begin{vmatrix}
        \hat{\varphi} & \hat{\chi} & \hat{\sigma} \\ \nonumber
       \sin \beta \cos \alpha & \sin \beta \sin \alpha& \cos \beta \\ \nonumber
         \frac{\dot{\phi}}{\dot{l}}V_{,l}-V_{,\phi} & \frac{\dot{\chi}}{\dot{l}}V_{,l}-V_{,\chi} & \frac{\dot{\sigma}}{\dot{l}}V_{,l}-V_{,\sigma}
    \end{vmatrix} \\ \nonumber
    &= \frac{1}{(\delta^{cd}V_{,c}V_{,d}-V_{,l}^2)^{\frac{1}{2}}} \times\\
    &\Big[\Big(V_{,\chi}\cos \beta - V_{,\sigma} \sin \beta \sin\alpha \Big)\hat{\varphi}
    +\Big(V_{,\sigma}\sin \beta \cos \alpha-V_{,\phi}\cos \beta\Big) \hat{\chi}
    +\Big(V_{,\phi}\sin \beta \sin \alpha - V_{,\chi}\sin \beta \cos \alpha\Big)\hat{\sigma}\Big].
\end{align}
Therefore,
\begin{align}
\nonumber
    \delta s_b &= \boldsymbol{\delta\Phi} \cdot \hat{b} =  \frac{1}{(\delta^{cd}V_{,c}V_{,d}-V_{,l}^2)^{\frac{1}{2}}}
    \begin{pmatrix}
       \delta \phi& \delta \chi& \delta \sigma
    \end{pmatrix}
     \begin{pmatrix}
      V_{,\chi}\cos \beta - V_{,\sigma} \sin \beta \sin\alpha\\
       V_{,\sigma}\sin \beta \cos \alpha-V_{,\phi}\cos \beta\\
    V_{,\phi}\sin \beta \sin \alpha - V_{,\chi}\sin \beta \cos \alpha
\end{pmatrix} \\ \nonumber
&=  \frac{1}{(\delta^{cd}V_{,c}V_{,d}-V_{,l}^2)^{\frac{1}{2}}} \times \\ &\Big[\Big(V_{,\chi}\cos \beta - V_{,\sigma} \sin \beta \sin\alpha\Big)\delta\phi+\Big( V_{,\sigma}\sin \beta \cos \alpha-V_{,\phi}\cos \beta\Big)\delta \chi+\Big(V_{,\phi}\sin \beta \sin \alpha - V_{,\chi}\sin \beta \cos \alpha\Big)\delta \sigma\Big].
\end{align}
Now, we have obtained the transformation matrix $\mathcal{M}_k$, which at each point gives us the perturbations in the kinematic basis from the field basis,
\begin{align}
     \begin{pmatrix}
        \delta l \\
        \delta s_n \\
        \delta s_l
    \end{pmatrix}= \mathcal{M}_k
    \begin{pmatrix}
        \delta \varphi \\
        \delta \chi \\
        \delta \sigma
    \end{pmatrix}.
\end{align}
Designating $(\delta^{cd}V_{,c}V_{,d}-V_{,l}^2)^{\frac{1}{2}}$ as $A$,  $\mathcal{M}_k$ will become
\begin{align}
    \mathcal{M}_k =
   \frac{1}{A} \begin{pmatrix}
        A\sin \beta \cos \alpha & A\sin \beta \sin \alpha & A\cos \beta \\
        V_{,l}\sin \beta \cos \alpha-V_{,\phi} & V_{,l}\sin \beta \sin \alpha-V_{,\chi} & V_{,l}\cos \beta-V_{,\sigma} \\
      V_{,\chi}\cos \beta - V_{,\sigma} \sin \beta \sin\alpha& V_{,\sigma}\sin \beta \cos \alpha-V_{,\phi}\cos \beta & V_{,\phi}\sin \beta \sin \alpha - V_{,\chi}\sin \beta \cos \alpha
    \end{pmatrix}.
\end{align}
Now, we have two transformation matrices, $\mathcal{M}_s$, which transforms the field perturbations from the field basis to the semikinematic basis, and $\mathcal{M}_k$, which transforms the field perturbations from the field basis to the kinematic basis. Using these matrices, we can build the transformation matrix from the semikinematic basis to the kinematic basis. We call this matrix $\mathcal{M}_{sk}$, which is
\begin{align}
    \mathcal{M}_{sk} = \mathcal{M}_{k} \mathcal{M}^{-1}_{s}.
\end{align}
Similarly, $\mathcal{M}^{-1}_{sk} = \mathcal{M}_{s}\mathcal{M}^{-1}_{k}$ transforms the kinematic basis to the semikinematic basis. Having these transformation matrices, we can calculate the isocurvature perturbation in the kinematic basis when needed.

\section{Deriving equations for $Q_l$, $\delta s_1$, and $\delta s_2$}\label{Appendix-B}
In order to derive the equations for $Q_l$, $\delta s_1$, and $\delta s_2$, first we write (\ref{diffequphibr}) - (\ref{diffequsigmabr}) in a matrix form
\begin{equation} \label{diffequsMatrix}
\frac{d^2}{dt^2}
\begin{pmatrix}
       Q_{\phi} \\
       Q_{\chi} \\
       Q_{\sigma}
\end{pmatrix}
+3H\frac{d}{dt}
\begin{pmatrix}
       Q_{\phi} \\
       Q_{\chi} \\
       Q_{\sigma}
\end{pmatrix}
+\mathcal{N}
\begin{pmatrix}
       Q_{\phi} \\
       Q_{\chi} \\
       Q_{\sigma}
\end{pmatrix}
=0,
\end{equation}
where
\begin{equation} \label{Ndef}
\mathcal{N}=
\begin{pmatrix}
       \frac{k^2}{a^2}+C_{\phi\phi} & C_{\phi\chi} & C_{\phi\sigma} \\
       C_{\chi\phi} & \frac{k^2}{a^2}+C_{\chi\chi} & C_{\chi\sigma} \\
       C_{\sigma\phi} & C_{\sigma\chi} & \frac{k^2}{a^2}+C_{\sigma\sigma}
\end{pmatrix}.
\end{equation}
Using the inverse transformation of \eqref{MSvstrans}, we can write \eqref{diffequsMatrix} as
\begin{equation} \label{diffequsMatrix2}
\frac{d^2}{dt^2}[\mathcal{M}^{-1}
\begin{pmatrix}
       Q_l \\
       \delta s_1\\
       \delta s_2
\end{pmatrix}]
+3H\frac{d}{dt}[\mathcal{M}^{-1}
\begin{pmatrix}
       Q_l \\
       \delta s_1\\
       \delta s_2
\end{pmatrix}]
+\mathcal{N}[\mathcal{M}^{-1}
\begin{pmatrix}
       Q_l \\
       \delta s_1\\
       \delta s_2
\end{pmatrix}] = 0\,,\\
\end{equation}
which becomes
\begin{equation}\label{diffequsMatrixf}
    \frac{d^2}{dt^2}
\begin{pmatrix}
       Q_l \\
       \delta s_1\\
       \delta s_2
\end{pmatrix}
+[2\mathcal{M}\frac{d}{dt}\mathcal{M}^{-1}+3H]\frac{d}{dt}
\begin{pmatrix}
       Q_l \\
       \delta s_1\\
       \delta s_2
\end{pmatrix}
+\mathcal{M}[\frac{d^2}{dt^2}\mathcal{M}^{-1}+3H\frac{d}{dt}\mathcal{M}^{-1}+\mathcal{N}\mathcal{M}^{-1}]
\begin{pmatrix}
       Q_l \\
       \delta s_1\\
       \delta s_2
\end{pmatrix} = 0\,.
\end{equation}
Using the following relations, which can be easily confirmed:
\begin{align}
\dot{\beta}&=\frac{V_{,s_2}}{\dot{l}}\,, \label{betadot}\\
\dot{\alpha}&= -\frac{V_{,s_1}}{\dot{l}\sin\beta}\,, \label{alphadot}\\
\cot\beta&= \frac{\dot{\sigma}}{\sqrt{\dot{\phi}^2+\dot{\chi}^2}} \\ \nonumber
\end{align}
and, by some long algebra, the coefficients of the second and the third terms of \eqref{diffequsMatrixf} can be worked out, 
\begin{equation} \label{AMatrix}
2\mathcal{M}\frac{d}{dt}\mathcal{M}^{-1}+3H =
\begin{pmatrix}
       3H & 2\frac{V_{,s_1}}{\dot{l}} & 2\frac{V_{,s_2}}{\dot{l}}\\
       -2\frac{V_{,s_1}}{\dot{l}} & 3H & 2\frac{V_{,s_1}}{\dot{l}}\cot\beta\\
       -2\frac{V_{,s_2}}{\dot{l}} & -2\frac{V_{,s_1}}{\dot{l}}\cot\beta & 3H
\end{pmatrix}
\end{equation}
and
\begin{equation}\label{BMatrix}
\mathcal{M}[\frac{d^2}{dt^2}\mathcal{M}^{-1}+3H\frac{d}{dt}\mathcal{M}^{-1}+\mathcal{N}\mathcal{M}^{-1}]=
\begin{pmatrix}
       \frac{k^2}{a^2}+C_{ll} & C_{ls_1} & C_{ls_2}\\
       C_{s_1l} & \frac{k^2}{a^2}+C_{s_1s_1} & C_{s_1s_2}\\
       C_{s_2l} & C_{s_2s_1} & \frac{k^2}{a^2}+C_{s_2s_2}
\end{pmatrix},
\end{equation}
therefore, we get \eqref{diffequQl}—\eqref{diffequS2}.








\bibliography{bibliography} 


\end{document}